\documentclass[11pt]{article}

\usepackage{arxiv}
\usepackage[utf8]{inputenc} 
\usepackage[T1]{fontenc}    
\usepackage{hyperref}       
\usepackage{url}            
\usepackage{booktabs}       
\usepackage{amsfonts}       
\usepackage{nicefrac}       
\usepackage{microtype}      
\usepackage{lipsum}
\usepackage{xurl}
\usepackage{amsmath,amssymb}
\usepackage{xcolor,colortbl}
\usepackage{mathtools}
\usepackage{bm}
\usepackage{graphicx}
\usepackage{pgfplots}
\usepackage{tikz}
\usepackage{layouts}
\usepackage[title]{appendix}
\usepackage{tabularx}
\usepackage{multirow}

\pgfplotsset{compat=newest}
\pgfplotsset{plot coordinates/math parser=false}
\newlength\figureheight
\newlength\figurewidth

\usepackage{stackengine}
\usepackage{soul}
\usepackage[ruled]{algorithm2e}

\renewcommand*{\vec}[1]{\mathbf{#1}}
\newcommand{\vect}[1]{\boldsymbol{\mathbf{#1}}}
\newcommand\xrowht[2][0]{\addstackgap[.5\dimexpr#2\relax]{\vphantom{#1}}}
\newcommand{\mathcolorbox}[2]{\colorbox{#1}{$\displaystyle #2$}}

\newcommand{\UR}[1]{{\color{red} #1}}

\newcommand\inputpgf[2]{{
\let\pgfimageWithoutPath\pgfimage
\renewcommand{\pgfimage}[2][]{\pgfimageWithoutPath[##1]{#1/##2}}
\input{#1/#2}
}}

\title{Bayesian Dynamical System Identification with Unified Sparsity Priors and Model Uncertainty}

\author{
  Prem Ratan Mohan Ram\\ 
  Institut für Dynamik und Schwingungen\\
  Technische Universität Braunschweig\\
  38106 Braunschweig, Germany \\
  \texttt{p.mohan-ram@tu-braunschweig.de} \\
   \And
  Ulrich Römer \\ 
  Institut für Dynamik und Schwingungen\\
  Technische Universität Braunschweig\\
  38106 Braunschweig, Germany \\
  \texttt{u.roemer@tu-braunschweig.de} \\
  \And
  Richard Semaan \\
  Institute of fluid mechanics\\
  Technische Universität Braunschweig\\
  38106 Braunschweig, Germany \\
  \texttt{r.semaan@tu-braunschweig.de} \\
}

\begin{document}
\maketitle

\begin{abstract}
This work is concerned with uncertainty quantification in dynamical system identification. 
Dynamical systems are ubiquitous in design and control applications and recent efforts focus on their data-driven construction. 
Our starting point is the sparse-identification of nonlinear dynamics (SINDy) framework, which reformulates system identification as a regression problem, where unknown functions are approximated from a sparse subset of an underlying library. 
In this manuscript, we formulate this system identification method in a Bayesian framework to handle parameter and structural model uncertainties. 
We present a general approach to enforce sparsity, which builds on the recently introduced class of neuronized priors. 
We perform comparisons between different variants such as Lasso, horseshoe, and spike and slab priors, which are all obtained by modifying a single activation function. 
We also outline how state observation noise can be incorporated with a probabilistic state-space model. 
The resulting Bayesian regression framework is robust and simple to implement. 
We apply the method to two generic numerical applications, the pendulum and the Lorenz system, 
and one aerodynamic application employing experimental measurements.
\end{abstract}

\section{Introduction}

Developing reliable dynamical models is crucial to all scientific disciplines ranging from epidemiology, neuroscience, finance, to turbulence. 
Dynamic modeling for the long-term features is a key enabler for physical understanding, state estimation from limited sensors signals, prediction, control, and optimization.
Dynamical system modeling has seen tremendous progress in the last decades, driven by algorithmic advances, accessibility to large data, and hardware speedups. 
One breakthrough in system identification was reported by Bongard and Lipson \cite{Bongard2007PNAS} using symbolic regression. 
The method performs a heuristic search of the best equation that describes the dynamics. 
Symbolic regression is however expensive and not easily scalable to large systems. 
This limitation may be bypassed by black-box techniques. 
These include Volterra series \cite{Fu1973IJC}, autoregressive models \cite{Chatfield2000Book} (e.g., ARX, ARMA, and NARMAX), eigensystem realization algorithm (ERA) \cite{Juang1994Book}, and neural network (NN) models \cite{Wang2015IEEE}. 
These approaches, however, have limited interpretability and provide little physical insights. Some (e.g. NN) require large volumes of data and long training time, luxuries that are not always at hand.

Recently, sparse identification of nonlinear dynamical systems (SINDy) was introduced \cite{brunton2016discovering}.
The approach relies on sparse regression for system identification from time-series data and shows remarkable performance. The regression allows the determination of the system coefficients, which multiply a set of candidate basis functions.
Sparsity is achieved through a regularization step that reduces complexity and the risk of overfitting.
However, both the regularization coefficient and the regularization norm are usually selected in a heuristic or ad hoc manner, without a general rigorous approach.
Moreover, the deterministic approach of SINDy does not consider the uncertainty in the inferred model parameters nor in the library of basis functions.

This shortcoming can be addressed by embedding the SINDy algorithm into a Bayesian framework, which is a subject of current great interest.
In \cite{niven2020bayesian}, the authors discuss the analogy between the SINDy formulation and a Bayesian MAP estimate. 
A Gaussian prior is used in a variational Bayesian approach to quantify uncertainties in the model. 
The approach directly employs derivative data, whereas the state is assumed to be perfectly known. 
A similar setting has been considered in \cite{zhang2018robust}, where a Bayesian thresholding algorithm was employed to obtain a SINDy variant with quantified uncertainty. The use of standard Gaussian priors in both cases simplifies the algorithmic treatment, however, it has limited capabilities in enforcing sparsity. Additionally, such a prior is mainly suited to handle parametric uncertainty in the dynamical model. 
Here, we emphasize the concept of model uncertainty, understood as the uncertainty in the structure of the identified model. When considering model uncertainty, a possible approach is to introduce a latent binary vector $\gamma_i \in \{0,1\}$, which indicates exclusion/inclusion of the associated basis function. 
Bayesian inference then allows to estimate the exclusion/inclusion probabilities $P(\gamma_i=0,1)$ together with the marginal distributions of the parameter. 
The subsequent prediction can then be based on model averaging, where multiple models are considered according to their probabilities. 
Another possibility uses a single model, for instance, the highest posterior probability model or the median model \cite{barbieri2004optimal}. 
The difficulty in using a model indicator vector is the exponential growth with the size of the library, which results in a complicated algorithmic treatment for complex models. 
This aspect of model uncertainty has been discussed in the context of the Bayesian Lasso in \cite{hans2010model}. 

In a recent contribution \cite{fuentes2021equation}, spike and slab priors, which are among the most efficient sparsity priors in Bayesian regression, have been introduced in the dynamical system learning context. 
The authors derive sparse models relying on median model selection and demonstrate an improved sparsity in the inferred model compared to the relevance vector machine. 
Here, again noise has been considered in the model equation, while the system states were assumed to be perfectly known. A nonlinear Bayesian learning approach, which is however not related to the SINDy framework, was put forth in \cite{sandhu2020nonlinear} employing automatic relevance determination priors.  Finally, a general formulation of Bayesian identification of dynamical system was recently presented in \cite{galioto2020bayesian}. 
There, the authors formulate a hidden Markov model of a time-discrete system, including process, observation and parameter uncertainty. 
The SINDy framework and dynamic mode decomposition are recovered as MAP estimates under specific choices for measurement and process noise, as well as the observation operator. 
Based on nonlinear Kalman approximations of the marginal Likelihood (marginalized over the inferred hidden states), the parameter posterior distribution is obtained with a dedicated MCMC algorithm.

In this study, we propose a flexible and automatable Bayesian framework for sparse dynamical system identification. 
The Bayesian formulation provides a rigorous approach for the choices of the residual and regularization terms in the sparse system regression framework. 
The method is enabled by the recently-proposed neuronized priors \cite{shin2021neuronized} that provide a unified formulation.
In particular, by choosing a suitable activation function, neuronized priors can represent highly efficient shrinkage priors such as discrete spike and slab, Lasso, and Horseshoe priors. 
Hence, we can explicitly target model selection and model uncertainty in addition to quantifying the uncertainty in model parameters. 
We outline how this flexible prior framework can be combined with the general setting of \cite{galioto2020bayesian}, which accounts for both process and observation noise. 
Considering process noise provides a natural means to account for numerical differentiation errors, which need to be taken into account if no state derivative data are available. 
Including observation noise, in turn, seems to be a natural, yet currently often omitted, step in the SINDy framework. 
Our approach is compared against the least-squares regression with thresholding and 
applied on two generic problems and one experimental dataset
using different shrinkage priors and different error minimization functions.
The proposed method is accurate and flexible.

The manuscript is structured as follows.
The original SINDy method is recalled in Section~\ref{sec:SINDy}. The proposed stochastic formulation with neuronized priors is introduced in Section~\ref{sec:Bsindy}
and applied on three different applications in Section~\ref{sec:numerics}.
The study is summarized in Section~\ref{sec:Summary}.

\section{Sparse system identification}
\label{sec:SINDy}

Before we introduce the Bayesian formulation, it is helpful to first review the deterministic SINDy algorithm.
Following \cite{brunton2016discovering}, we consider dynamical systems of the form 
\[
\dot{\vec{x}} = \frac{d \vec{x}}{dt} = \vec{f}(\vec{x}(t)), \quad \vec{x}(0) = \vec{x}_0,
\]
where $\vec{x} \in \mathbb{R}^{n}$ refers to the state of the system with initial condition $\vec{x}_0$. Our aim is to learn the function $\vec{f}: \mathbb{R}^n \rightarrow \mathbb{R}^n$ from time-series data assuming a set of basis functions. 
The SINDy paradigm formulates this learning task as a parameter estimation problem. 
Without loss of generality, let us introduce the problem setting with a simple one-dimensional state $x$. 
Possible library function candidates for $\vec{f}$ could be a linear and quadratic basis, hence, 
\begin{equation}
    \label{eq:example_ode}
    \dot{x}(t) \approx \xi_1 x(t) + \xi_2 x^2(t), \quad x(0) = x_0.
\end{equation}
The parameters $\vect{\xi} = (\xi_1,\xi_2)^\top$ are estimated as the solution of a regression problem. 
In vector notation, \eqref{eq:example_ode} can be expressed as
\begin{equation}
    \dot{\vec{x}} \approx \underbrace{[\vec{x} \ \vec{x}^2]}_{=\Theta(\vec{x})} \vect{\xi}\,,
\end{equation}
where the basis functions $\vec{x} = (x(t_1),\ldots,x(t_m))^\top$ and $\vec{x}^2 = (x^2(t_1),\ldots,x^2(t_m))^\top$ are vectors evaluated over discrete time $t_i, i=1,\ldots,m$. 
The regression problem reads 
\begin{align}
    \vect{\xi}^* &= \mathrm{argmin}_{\vect{\xi}} \sum_{i=1}^m (\dot{x}(t_i) - \xi_1 x(t_i) - \xi_2 x^2(t_i))^2 \\
    &= \mathrm{argmin}_{\vect{\xi}} \| \dot{\vec{x}} - \Theta(\vec{x}) \vect{\xi} \|^2_2,
\end{align}
where $\| \cdot\|_2$ refers to the Eucledian norm. 
Crucial for the success of this approach is the choice of a library of basis functions. 
In particular, we assume that the basis library is sufficiently rich to capture all relevant dynamics. 
Note that since the derivative is rarely available, a common approach employs a numerical approximation, i.e., a difference quotient or total variation denoising. After identification, the dynamics of each component can be expressed as 
\begin{equation}
    \label{eq:SINDy_component}
    \dot{x} \approx \Theta(x(t)) \vect{\xi}, 
\end{equation}
where $\Theta(x(t))$ contains the library of functions, as given for instance in \eqref{eq:example_ode}. 
To keep the notation simple, we maintain the case of a one-dimensional state variable for the time being and provide details on the general case only later. 
We note that also in the general case with higher state dimensions, each elements of the vector $\vec{f}$ are learned independently, resulting in an individual parameter vector $\vect{\xi}$ for each dynamical equation. 

So far, we have assumed that both the state and the time derivative of the state are known, which is uncommon in real-life applications. 
With a suitable approximation, such as a forward Euler different quotient, we can rewrite \eqref{eq:SINDy_component} as
\begin{equation}
    \label{eq:SINDy_discrete}
    \frac{x(t_{l}) - x(t_{l-1})}{\Delta t} \approx \Theta(x(t_{l-1})) \vect{\xi}, \quad l = 2,\ldots,m,
\end{equation}
where we have assumed a uniform time step size $\Delta t$, for simplicity. 
The regression problem is now expressed as,
\begin{equation}
\vect{\xi}^* = \mathrm{argmin}_{\vect{\xi}} \sum_{l=2}^m \left(\frac{x(t_{l}) - x(t_{l-1})}{\Delta t} - \Theta(x(t_{l-1})) \vect{\xi} \right)^2.
\end{equation}
This approach assumes that the state is perfectly known, i.e., observation error is neglected, which is a potential source of bias. Consider for instance a data model of the form 
\begin{equation}
    \label{eq:error_in_measurement}
    y(t) = x(t) + e(t),
\end{equation}
which expresses that the true state $x$ cannot be observed, instead we only have access to the surrogate (sometimes referred to as the observable) $y$. 
In a general parameter estimation setting, it has been reported \cite{liang2008parameter} that employing $y$ instead of $x$ results in a biased estimate. 
A possible remedy is using a pseudo-least-squares estimate, where first estimators of the solution and the derivative are introduced as $\hat{x},d \hat{x}/dt$, which are then used to solve the regression problem. The properties of the pseudo-least squares estimator, as well as the connection to error in measurement models have been investigated in \cite{liang2008parameter}. 

We proceed by introducing a matrix vector version of \eqref{eq:SINDy_discrete}, by setting $z_{l-1}= (x(t_{l}) - x(t_{l-1}))/\Delta t$, $\vec{z} = (z_1,\ldots,z_{m-1})^\top$ and the $\mathbb{R}^{m-1\times p}$ matrix $\vec{D}$, where the row $j$ of $\vec{D}$ is given by $\Theta(\vec{x}(t_{j}))$. Then, the regression model can be expressed as
\begin{equation}
    \label{eq:linear}
    \vec{z} \approx \vec{D} \vect{\xi},
\end{equation}
which is a standard linear model. One main assumption of the SINDy approach is the sparsity of the parameter vector $\vect{\xi}$. 
Sparsity is often present in mathematical models of physical systems, which are frequently amenable to reduced order and low-rank modeling. 
Specifically, sparsity refers to the fact that many coefficients of $\vect{\xi}$ are zero. 
Hence, a dedicated (linear) regression approach needs to incorporate some regularization technique. 
A regularized least-squares approach to solve \eqref{eq:linear} minimizes the objective function 
\begin{equation}
    \label{eq:SINDy_regularized}
    \vect{\xi}^* = \mathrm{argmin}_{\vect{\xi}} \|\vec{z} - \vec{D}\vect{\xi}\|_2^2 + \lambda^2 \|\vect{\xi}\|^2.
\end{equation}
Different choices are available concerning the norm of the regularization term $\|\vect{\xi}\|$; 
popular choices include the $l^2$-norm (ridge regression), the $l^1$-norm (LASSO regression) and the $l^0$-norm (counting the number of non-zero elements). 
The choice of a Lasso least-square approach has already been suggested in the original SINDy paper \cite{brunton2016discovering}. 
However, eventually, a standard least-square algorithm together with iterative thresholding was recommended. 
The SINDy procedure can be compactly summarized as in Algorithm~\ref{alg:SINDY} \cite{zhang2019convergence}.
\begin{center}
\begin{minipage}{.6\linewidth}
\begin{algorithm}[H]
\SetAlgoLined
\KwResult{Approximation of $\vect{\xi}^*$}
 $\vect{\xi}^{(0)} = \vec{D}^\dagger \vec{z}$ $\quad$ (pseudoinverse $\vec{D}^\dagger$)\;
 $S= \{1,\ldots,p\}$ \;
 \For{$k=1,\ldots,k_\text{max}$}{
    \If{$|\xi_j^{(k-1)}| \geq \lambda, j=1,\ldots,p,$}{
  Set $S= S \setminus \{j\}$\;
 }
 $\vect{\xi}^{(k)} = \mathrm{argmin}_{\vect{\xi}_S} \|\vec{D} \vect{\xi}_S - \vec{z} \| \quad (\vect{\xi}_{S,i} = 0, \forall i \notin S)$ \;
 }
 \caption{SINDy}
 \label{alg:SINDY}
\end{algorithm}
\end{minipage}
\end{center}
%
%
Hence, in the first step, an ordinary least-squares problem is solved. 
It can be shown \cite{zhang2019convergence} that this iterative thresholding converges to a local minimum of \eqref{eq:SINDy_regularized}. 
Hence, the minimization can serve as a common starting point for a Bayesian SINDy formulation.

Before moving to the Bayesian setting, we briefly discuss how to handle the general case, with multiple state variables. To this end, we introduce the matrices
\begin{align*}
    \vec{X} = 
        \left (
        \begin{array}{c}
             \vec{x}^\top(t_1)  \\
             \vec{x}^\top(t_2)  \\
             \vdots  \\
             \vec{x}^\top(t_m)
        \end{array}
        \right ),
     \qquad \dot{\vec{X}} = 
        \left (
        \begin{array}{c}
             \dot{\vec{x}}^\top(t_1)  \\
             \dot{\vec{x}}^\top(t_2)  \\
             \vdots  \\
             \dot{\vec{x}}^\top(t_m)
        \end{array}
        \right ), \qquad \vec{X}, \dot{\vec{X}} \in \mathbb{R}^{m \times n}.
\end{align*}
The library for approximating the unknown function $\vec{f}$ is then constructed as 
\begin{equation}
    \Theta(\vec{X}) = \left [ \vec{1}  \quad \vec{X} \quad \vec{X}^2  \cdots \right ], \quad \Theta(\vec{X}) \in \mathbb{R}^{m \times p},
\end{equation}
where, for instance, row number $i$ of $\vec{X}^2$ is given by the vector representation of $\vec{x}(t_i) \otimes \vec{x}(t_i)$. With this notation at hand, we seek for an approximation 
\begin{equation}
    \label{eq:SINDy}
    \dot{\vec{X}} \approx \Theta(\vec{X}) \vect{\xi},
\end{equation}
with coefficients $\vect{\xi} \in \mathbb{R}^{p \times n}$, to be determined. 

\section{Bayesian SINDy with neuronized priors}
\label{sec:Bsindy}
To incorporate measurement noise and uncertainty, the linear model \eqref{eq:linear} is modified as 
\begin{equation}
\label{eq:linearwNoise}
\vec{z}=\vec{D}\vect{\xi} + \vect{\eta},
\end{equation} 
where $\vec{D}$ represents the design matrix. 
We assume, for the time being, perfect knowledge of the state and hence of $\vec{D}$ and consider uncertainty only in the data derivative. 
A Bayesian approach first formulates a prior density $p(\vect{\xi})$ for the unknown model parameter vector. The likelihood expresses the probability of obtaining the data $\vec{z}$ as
\begin{equation}
    \label{eq:likelihood}
    p(\vec{z}|\vect{\xi}) \sim \mathcal{N}(\vec{D}\vect{\xi},\vect{\Sigma}_{\vect{\epsilon}}),
\end{equation}
where $\vect{\Sigma}_{\vect{\epsilon}}$ denotes the covariance matrix of the Gaussian noise $\vect{\eta}$. Bayes' theorem then expresses the posterior density as 
\begin{equation}
    \label{eq:Bayes}
    p(\vect{\xi}|\vec{z}) \propto p(\vec{z}|\vect{\xi}) p (\vect{\xi}).
\end{equation}
From \eqref{eq:Bayes}, a MAP estimate is obtained by solving 
\begin{equation}
    \vect{\xi}^* = \mathrm{argmin}_{\vect{\xi}} \left(- \log \left(p(\vect{\xi}|\vec{z})\right) \right)= \mathrm{argmin}_{\vect{\xi}} \left ( \|\vec{z} - \vec{D}\vect{\xi}\|_{\vect{\Sigma}_{\vect{\epsilon}}^{-1}}^2 - \log \left( p(\vect{\xi}) \right ) \right),
\end{equation}
with the $\vect{\Sigma}_{\vect{\epsilon}}^{-1}$-weighted $l^2$-norm $\| \cdot \|_{\vect{\Sigma}_{\vect{\epsilon}}^{-1}}$. 
A comparable formulation to \eqref{eq:SINDy_regularized} is obtained by choosing a Laplace prior $p(\vect{\xi}) \propto \exp (- \lambda \|\vect{\xi}\|_1)$, whereas a normal prior yields a ridge-regression. 
Many more sparsity-inducing priors can be found in the literature. 
Examples include the horseshoe prior and the spike and slab priors. 
The latter results in a MAP estimates with tight connections to $l^0$-regularization, which completes the analogy between Bayesian and standard sparse regression \cite{polson2019bayesian}.

A more general approach employs the probabilistic state space model formulation 
\begin{align}
    x_{i+1} &= x_i + \Delta t \Theta(x_i) \vect{\xi} + \eta_i, &\eta_i \sim \mathcal{N}(0,\sigma_\eta^2), \label{eq:model_process}\\
    y_j &= x_j + \varepsilon_j,  &\varepsilon_j \sim \mathcal{N}(0,\sigma_\varepsilon^2)\label{eq:model_observation},
\end{align}
where $i=1,\ldots,m-1$ and $j=1,\ldots,k$. 
The measurements can be either acquired at every discrete time step ($k=m$), or at subsets.
The assumption of independent, zero mean Gaussian observation noise is quite common and often justifiable. The process noise $\eta_i$ can represent the discretization error associated with the numerical derivative in our case. In this case, the Gaussian assumption is more debatable since the numerical error is typically structured. 
We maintain the independent normality assumption nevertheless and refer to \cite{conrad2017statistical} for treatment in this direction. 
Recent studies on Bayesian generalizations of SINDy consider $\eta_i$, but ignore $\varepsilon_j$. 
In fact, the standard deterministic SINDy formulation is obtained as a MAP estimate, when setting $\sigma_\eta=1$ and $\sigma_\varepsilon=0$, see Theorem 4.2 of \cite{galioto2020bayesian}. 

Following \cite{sarkka2013bayesian}, we proceed by recasting \eqref{eq:model_process} and \eqref{eq:model_observation} as 
\begin{align}
\vect{\xi} &\sim p(\vect{\xi}), &\text{(prior)},\\
x_i &\sim p(x_i|x_{i-1},\vect{\xi}), &\text{(dynamics)}, \\
y_j &\sim p(y_j|x_{j}), &\text{(observation)},
\end{align}
where we assume perfect knowledge of the initial conditions, for simplicity. We now collect the measurements in a vector $\vec{y} = (y_1,\ldots,y_k)^\top$. Accounting for observation noise now requires to jointly estimate $(\vec{x},\vect{\xi})$. Here, we are mainly interested in learning the dynamics, encoded in $\vect{\xi}$, which is guided by the marginal posterior density 
\[
p(\vect{\xi}|\vec{y}) = \int p(\vec{x},\vect{\xi}|\vec{y}) \ \mathrm{d} \vec{x}.
\]
Computing the marginal posterior, which is mainly based on a formula for the marginal likelihood $p(\vec{y}|\vect{\xi})$, is commonly carried out recursively, including a prediction and an update step, exploiting the Markovian structure of the dynamical model. 
The main motivations being computational simplicity and the sequential character of new measurements. This has been covered in \cite{galioto2020bayesian} and is not presented in any detail here. 

Our goal is to obtain a quantification of uncertainty in the inferred model parameters and models, where the latter is tightly related to the inclusion probabilities of the individual library functions. 
Hence, we have to go beyond MAP estimates and consider the posterior distribution. 
There exists a large body of literature addressing the challenges in sparse Bayesian regression. 
Here, we recall two main classes, continuous global-local and spike and slab priors. 
A continuous global-local prior is given as 
\begin{equation}
    \label{eq:global-local}
    p(\xi_j|\lambda_j,\tau) \sim \mathcal{N}(0,\tau \lambda_j^2),
\end{equation}
where $\tau,\lambda_j$ are called the global and local shrinkage parameter or variance components, respectively. 
This type of model is discussed in detail, for instance in \cite{polson2010shrink}. 
For simplicity, we consider a zero mean in the discussion of different priors. 
The global shrinkage parameter $\tau$ pushes the coefficients towards zero, whereas some parameters are allowed to assume large values if needed, by choosing a heavy-tailed distribution for $\lambda_j$. 
The model \eqref{eq:global-local} is hierarchical and different choices for the distributions of $\tau,\lambda$ give rise to different priors. 
Popular choices are the Horseshoe prior \cite{carvalho2010horseshoe} or the Bayesian Lasso \cite{park2008bayesian}. 

On the other hand, spike and slab priors employ a mixture distribution
\begin{equation}
    \label{eq:spike_and_slab_continuous}
    p(\xi_j|c,\delta,\gamma_j) \sim \gamma_j \mathcal{N}(0,\tau c^2) + (1-\gamma_j) \mathcal{N}(0,\tau \delta^2),
\end{equation}
where the Bernoulli variable $\gamma_j$ expresses the inclusion probability of the coefficient $\xi_j$. By setting $\delta \ll c$ the second distribution on the right-hand-side of \eqref{eq:spike_and_slab_continuous} is concentrated around zero (spike), whereas $\mathcal{N}(0,\tau c^2)$ is less concentrated (slab). 
Spike and slab priors often provide excellent sparsity in regression \cite{piironen2017sparsity}. 
Whereas \eqref{eq:spike_and_slab_continuous} describes the continuous version, a discrete counterpart is obtained by replacing $\mathcal{N}(0,\tau \delta^2)$ with a delta function at the origin as
\begin{equation}
    \label{eq:spike_and_slab_discrete}
    p(\xi_j|c,\delta,\gamma_j) \sim \gamma_j \mathcal{N}(0,\tau c^2) + (1-\gamma_j) \delta_0.
\end{equation}

\begin{figure}[t!]
    \begin{minipage}{0.33\textwidth}
    \centering
    \begin{tikzpicture}
    \node at (0,2) {Lasso};
    \node at (0,0) {\includegraphics{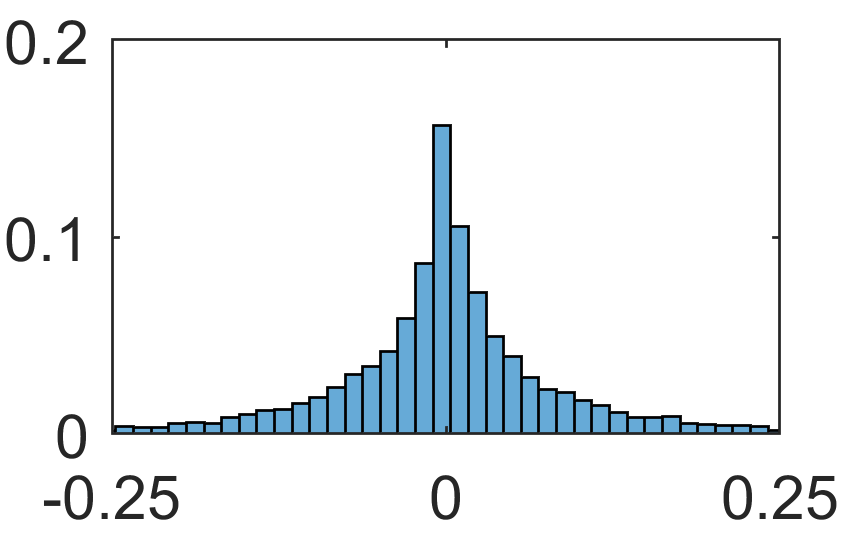}};
    \node at (0,-2) {$\xi$};
    \end{tikzpicture}
    \end{minipage}
    \begin{minipage}{0.33\textwidth}
    \centering
    \begin{tikzpicture}
    \node at (0,2) {Horseshoe};
    \node at (0,0) {\includegraphics{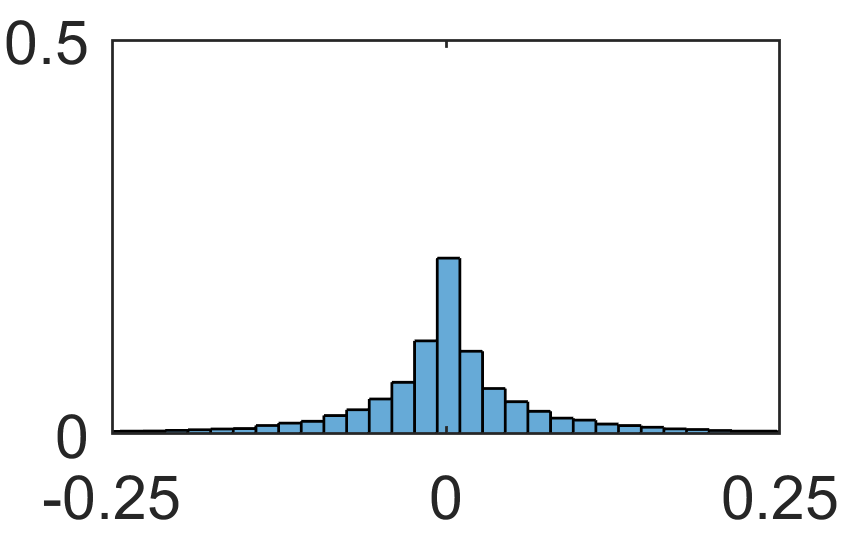}};
    \node at (0,-2) {$\xi$};
    \end{tikzpicture}
    \end{minipage}
    \begin{minipage}{0.33\textwidth}
    \centering
    \begin{tikzpicture}
    \node at (0,2) {ReLu};
    \node at (0,0) {\includegraphics{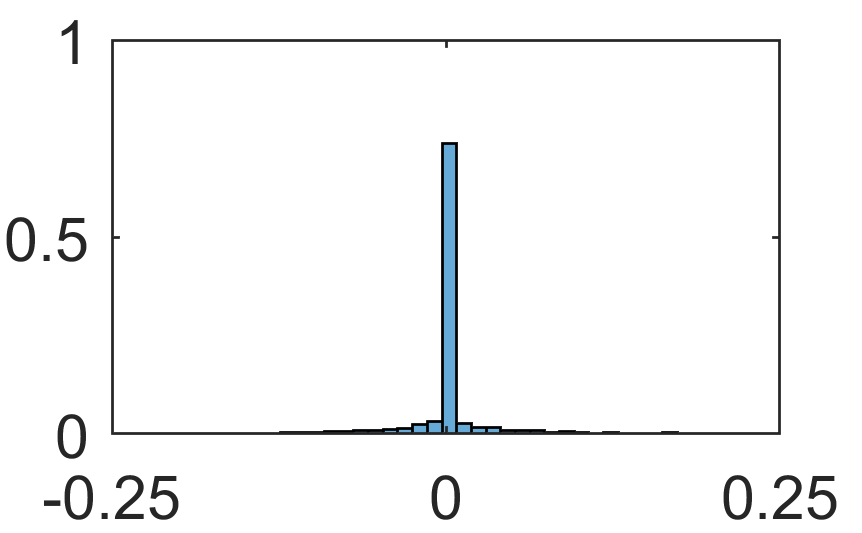}};
    \node at (0,-2) {$\xi$};
    \end{tikzpicture}
    \end{minipage}
    \caption{Examples of neuronized priors. All priors enforce sparsity by allocating probability mass near (or at) zero. Left: Lasso prior for $\tau_w = 0.1$, which is obtained by setting $T(\xi) = \xi$, $\alpha_0=0$.  Middle: Approximate horseshoe prior with $T(\xi) = \exp(0.37 \mathrm{sign}(\xi)\xi^2 + 0.89\xi + 0.08)$, $\alpha_0 = 0$ and $\tau_w = 0.05$. The distribution is only asymptotically equivalent to the horseshoe prior. Right: Discrete spike and slab prior with $T(\xi) = \max(0,\xi)$, which corresponds to the ReLu activation function. 
    In this case $\alpha_0 = 0.5$ plays the role of the sparsity parameter and the slab variance is chosen as $\tau_w=0.1$. The vertical axes-range differs among the three figures to allow for a better visualization.}
    \label{fig:neuroPrior}
\end{figure}
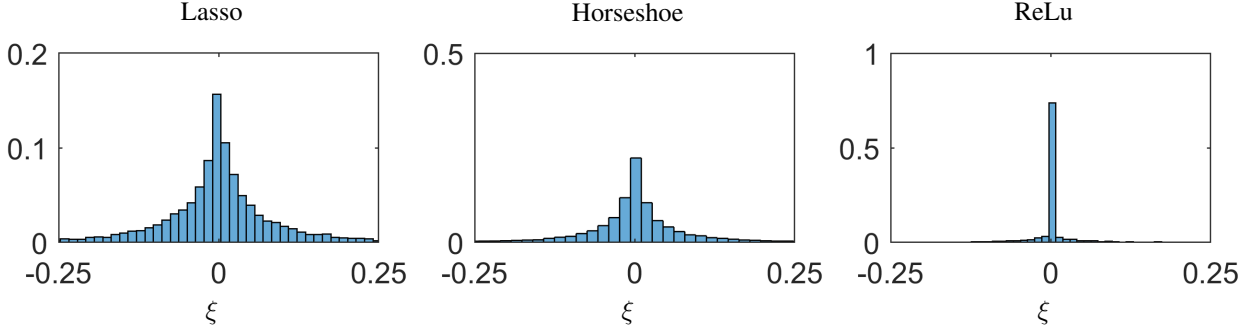

In this study, we combine multiple state-of-the-art techniques for dynamical system identification with uncertainty quantification. 
To this end, we employ a recently proposed formulation of priors, the so-called neuronized priors \cite{shin2021neuronized}, which provide a unified treatment. 
Moreover, we utilize an efficient MCMC algorithm \cite{shin2021neuronized} to infer posterior distributions of the model parameters. 
A neuronized prior is formulated as, 
\begin{align}
    \label{eq:neuronized}
    p(\xi_j|\alpha_0) \sim T(\alpha_j - \alpha_0) w_j, 
\end{align}
where $\alpha_0$ is a fixed hyperparameter and $T$ is an activation function such as the ReLu function. 
The notion of an activation function is borrowed from artificial neural networks, whence the name neuronized prior. The model is hierarchical, i.e., 
\begin{align}
\label{eq:neuronized_hierarchy}
    p(\alpha_j) &\sim N(0,1), \\
    p(w_j|\tau_w) &\sim N(0,\tau_w^2).    
\end{align}
The main interest of the formulation is its generality, i.e., by choosing different activation functions, a spike and slab, Horseshoe, and Lasso prior are recovered, at least asymptotically. 
Additionally, the formulation allows obtaining efficient algorithms, avoiding the necessity to use latent binary indicator variables, as pointed out in \cite{nie2020bayesian}. 
Different versions of neuronized priors, together with their descriptions are presented in Figure~\ref{fig:neuroPrior}.


Having formulated the prior, two different strategies can be found in the literature for model selection. On the one hand, variable selection methods aim to identify a single model without necessarily quantifying uncertainty. 
We mention for instance the expectation-maximization algorithm presented in \cite{rovckova2014emvs}. 
This path is typically chosen, if the sampling from the full posterior distribution is too expensive.
Since the problems we consider here, in particular the size of the data and the library, are not too large, we employ an MCMC approach instead. 
The algorithm to approximate the marginal posterior distribution $p(\vect{\alpha},\vect{w}|\vec{y},\tau_w,\alpha_0)$ is summarized in Algorithm~\ref{alg:BSindy}, which is adopted from \cite{shin2021neuronized}. This algorithm successively updates $\vec{w}$ and $\vec{\alpha}$, where  $\vec{w}$ can be sampled from a normal distribution with covariance matrix $$\tilde{\vect{\Sigma}}(\vect{\alpha})= \sigma_{\eta}^2 (D_{\vect{\alpha}} \vec{D}^\top \vec{D}D_{\vect{\alpha}} + \sigma_\eta^2 \tau_w^{-2} \vec{I}_{p \times p})^{-1}, \ D_{\vect{\alpha}}= \mathrm{diag}(\alpha_1,\ldots,\alpha_p)$$ and mean value $\tilde{\vect{\mu}}(\vec{y},\vect{\alpha}) = \tilde{\vect{\Sigma}}(\vect{\alpha}) D_{\vect{\alpha}} \vec{D}^\top \vec{y}$.  The sampling distribution for $\vec{\alpha}$ depends on the chosen activation function and is therefore, not explicitly stated. 

Based on the MCMC approximation, we obtain the posterior of the state as
\begin{equation}
    \label{eq:postrior_predictive}
    p(x(t)|\vec{y},\tau_w,\alpha_0) = \int  p(x(t)|\vec{w},\vect{\alpha},\tau_w,\alpha_0)  p(\vec{w},\vect{\alpha}|\vec{y},\tau_w,\alpha_0)  \ \mathrm{d} \vec{\vect{\alpha}} \mathrm{d} \vec{\vec{w}}.
\end{equation}
We will comment on the choice of the remaining hyperparameters $\tau_w,\alpha_0$ in the numerical results section. 
As previously mentioned, estimating the marginal Likelihood requires a Kalman filter together with marginalization over the state. In the simple case where $\sigma_\varepsilon = 0$, this is not required and we obtain the simplified expression 
\[
p(\vec{y}|\vect{\alpha},\vect{w}) \propto \mathrm{exp}\left( - \frac{1}{2} \frac{\|\vec{D} \vect{\xi}(\vect{\alpha},\vect{w}) - \vec{z} \|_2^2}{\sigma_\eta^2} \right),
\]
which will be used in the following examples.
\begin{center}
\begin{minipage}{.6\linewidth}
\begin{algorithm}[H]
\SetAlgoLined
\KwIn{Hyper parameter $\tau_w,\alpha_0$ $\qquad$ (omitted in formulas below)} 
\KwIn{Estimate of data marginal Likelihood  $\hat{p}(\vec{y}|\vect{\alpha},\vect{w}) \approx p(\vec{y}|\vect{\alpha},\vect{w})$}
\KwResult{Sample $(\vec{w}^{(i)},\vect{\alpha}^{(i)})$ of posterior density $p(\vect{\alpha},\vect{w}|\vec{y})$}
\For{$i=1,\ldots,N$}{
Sample $\vec{w}^{(i)}$ from $\vec{w}|\vec{y},\vect{\alpha} \sim \mathcal{N}(\tilde{\boldsymbol{\mu}}(\vec{y},\vect{\alpha}),\tilde{\boldsymbol{\Sigma}}(\vect{\alpha}))$\;
Sample $\vect{\alpha}^{(i)}$ based on $\hat{p}(\vect{\alpha}|\vec{y},\vect{w}^{(i)})$ with a random walk Metropolis Hastings algorithm\;
}
 \caption{MCMC with neuronised prior}
 \label{alg:BSindy}
\end{algorithm}
\end{minipage}
\end{center}


\section{Results}
\label{sec:numerics}

The Bayesian SINDy algorithm with neuronised MCMC is applied on three systems; two generic examples, the linear pendulum and the Lorenz system, and one aerodynamic application, for which experimental measurement data are used.

\subsection{Linear Pendulum}
\label{sec:pendulum}

\begin{figure}[t]
\begin{minipage}{0.49\textwidth}
\centering
\begin{tikzpicture}
\node at (0,0) {\includegraphics{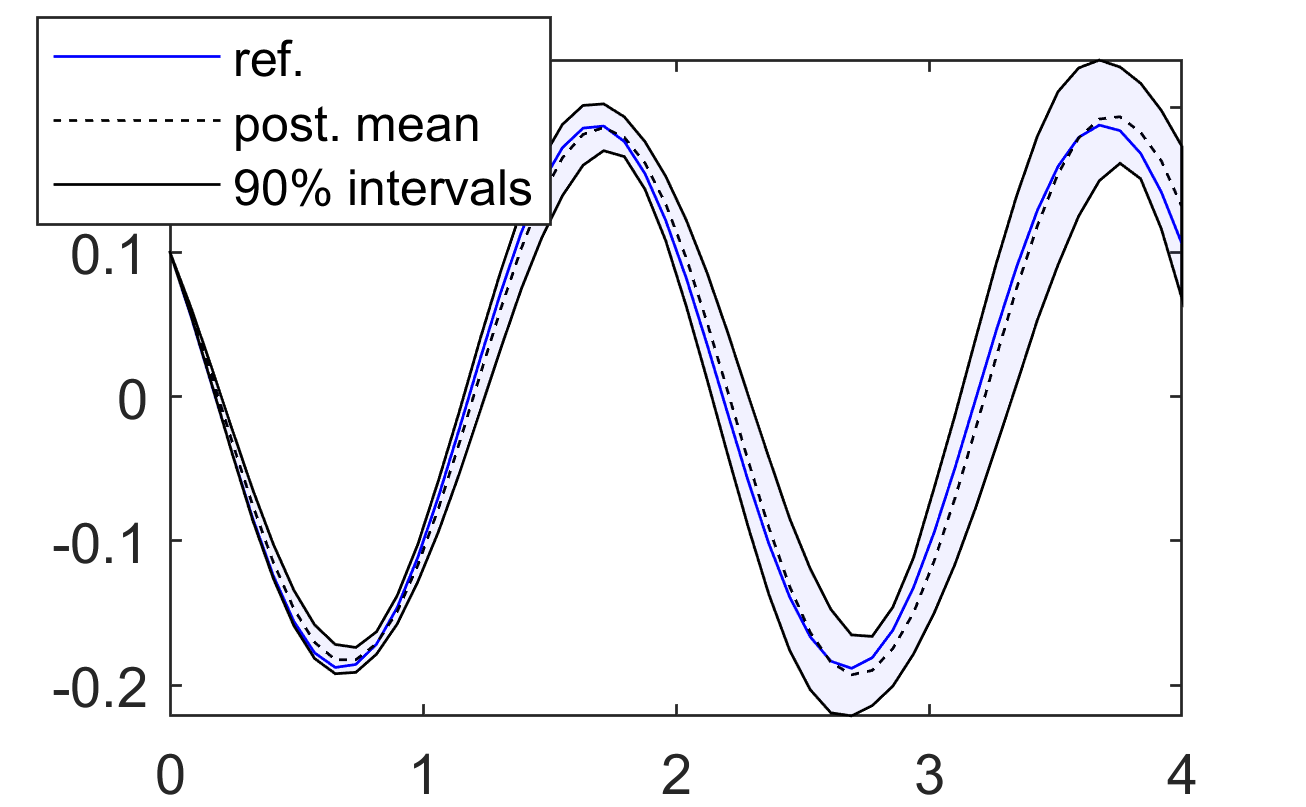}};
\node at (-4.2,0) {$x_1$};
\node at (0,-3) {$t$};
\end{tikzpicture}
\end{minipage}
\begin{minipage}{0.49\textwidth}
\centering
\begin{tikzpicture}
\node at (0,0) {\includegraphics{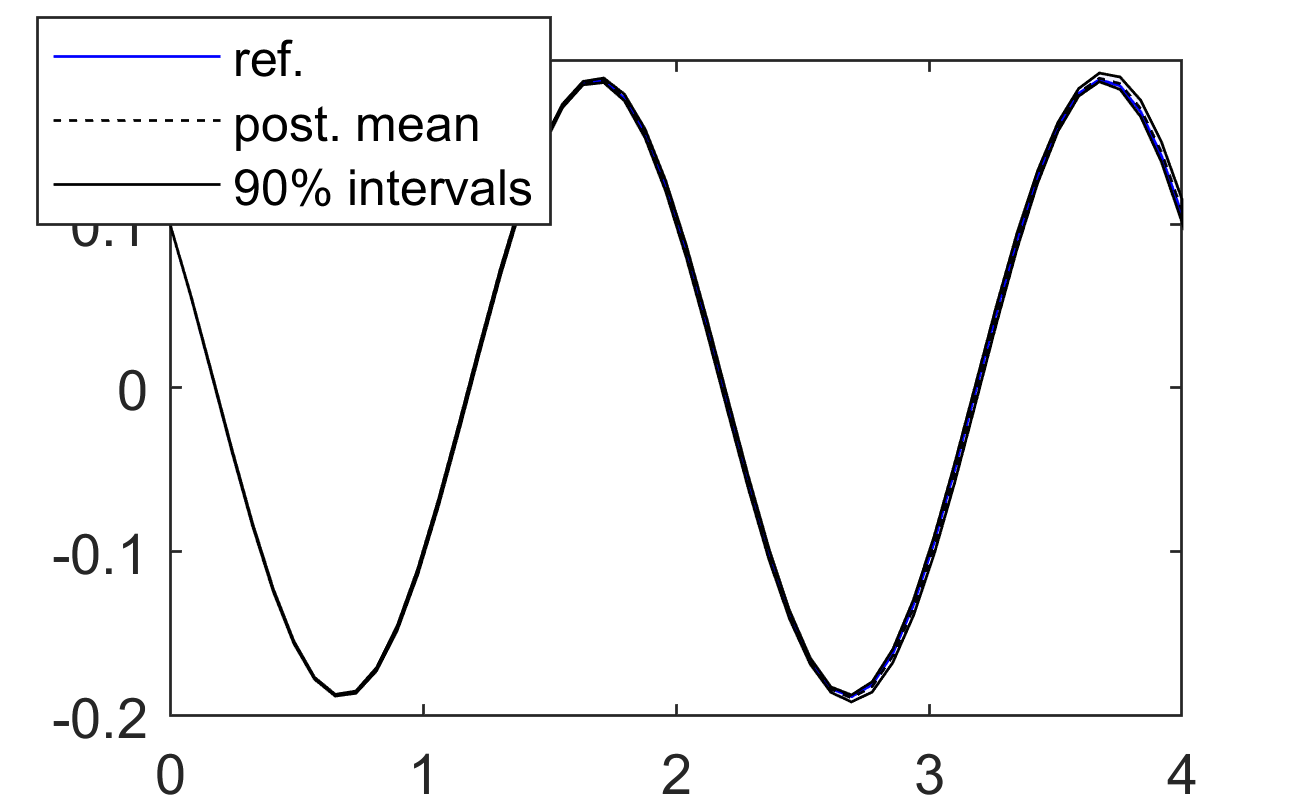}};
\node at (-4.2,0) {$x_1$};
\node at (0,-3) {$t$};
\end{tikzpicture}
\end{minipage}
\caption{Results for the identification of the linear pendulum with a Lasso prior. Reference trajectory, posterior mean and credible intervals covering $90 \%$ of the posterior samples for noise level $\sigma_\eta = 10^{-1}$ (left), and noise level $\sigma_\eta = 10^{-2}$ (right).}
\label{fig:pendulum_Lasso_posterior}
\end{figure}
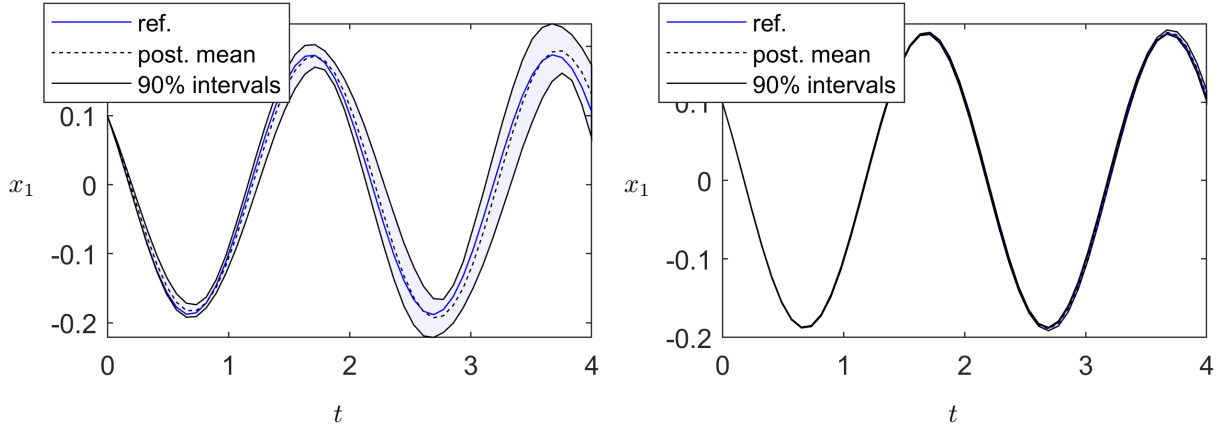
\begin{figure}[t]
\begin{minipage}{0.49\textwidth}
\centering
\begin{tikzpicture}
\node at (0,0) {\includegraphics{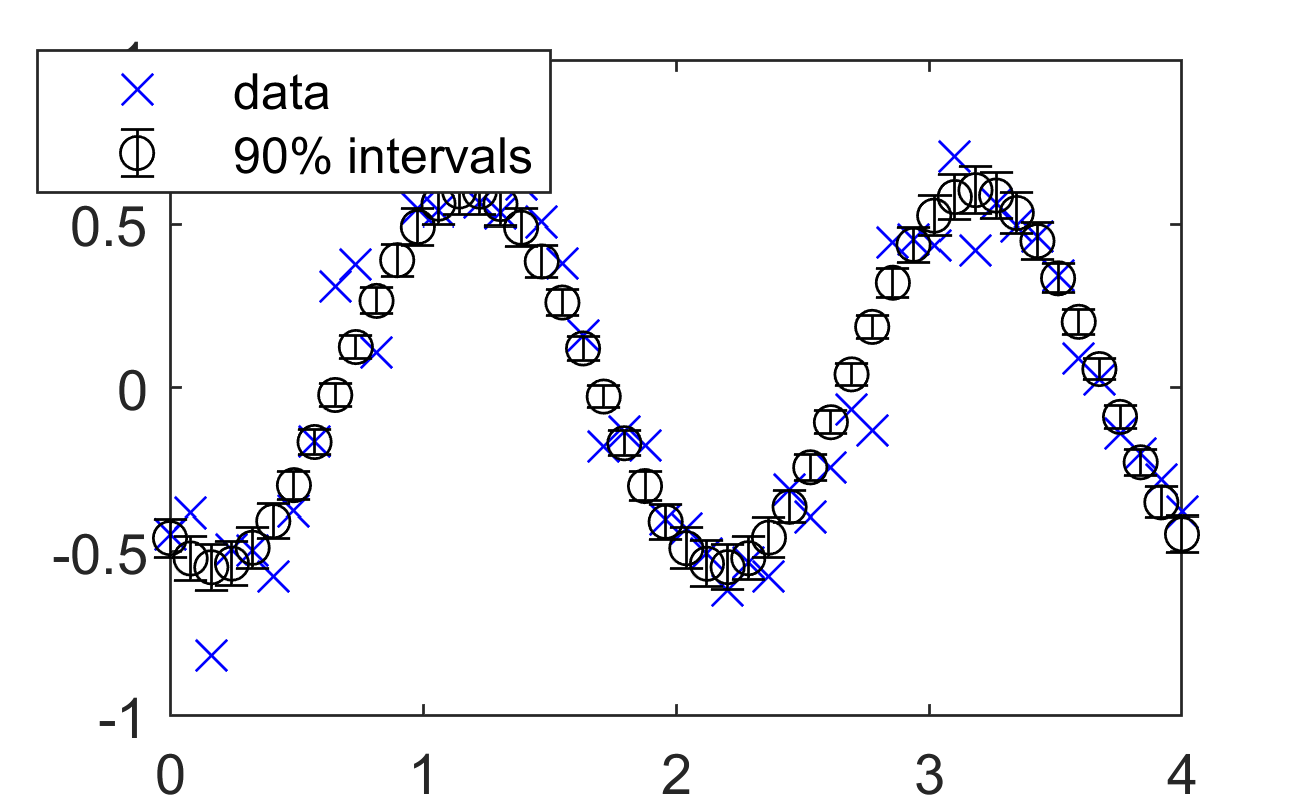}};
\node at (-4.2,0) {$\dot{x}_1$};
\node at (0,-3) {$t$};
\end{tikzpicture}
\end{minipage}
\begin{minipage}{0.49\textwidth}
\centering
\begin{tikzpicture}
\node at (0,0) {\includegraphics{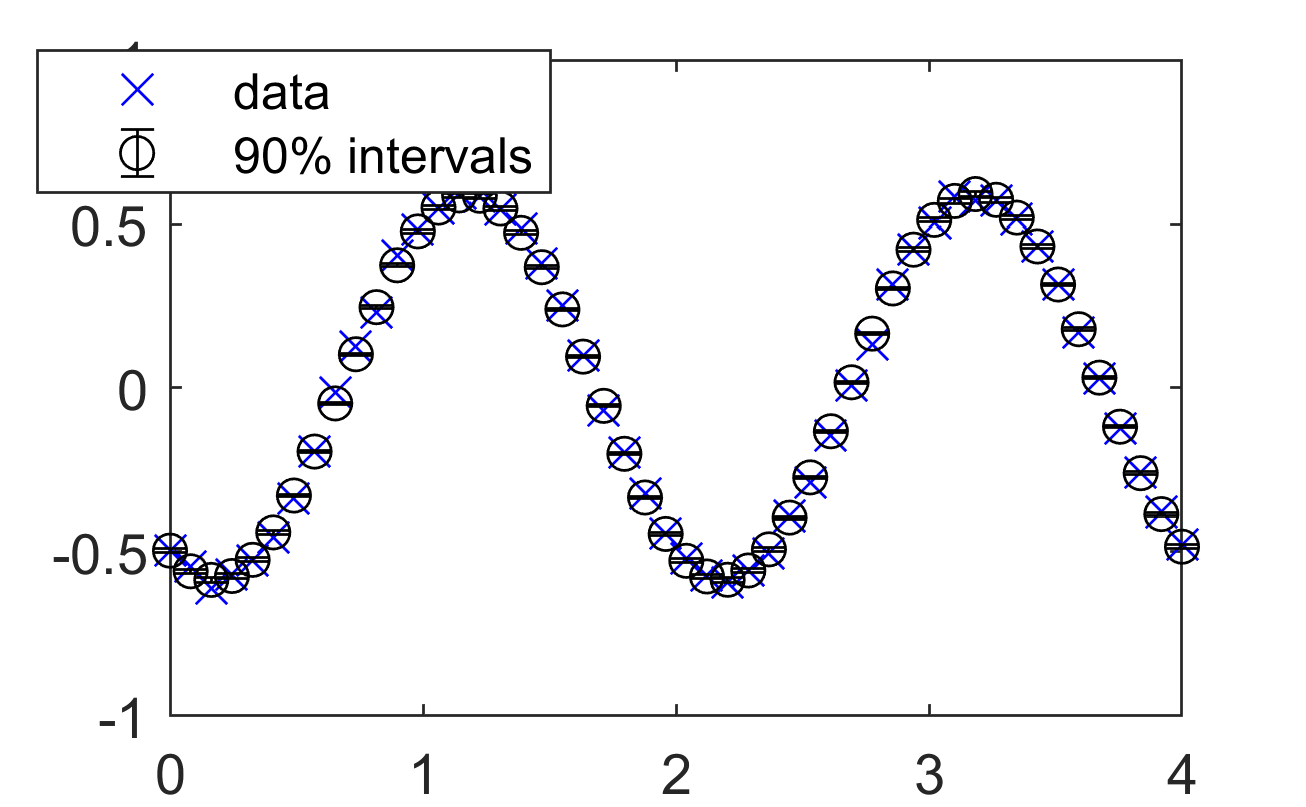}};
\node at (-4.2,0) {$\dot{x}_1$};
\node at (0,-3) {$t$};
\end{tikzpicture}
\end{minipage}
\caption{Results for the identification of the linear pendulum with a Lasso prior. Time derivative data $\vec{z}$ and posterior of $\Theta(\vec{X}) \vect{\xi}$. The $90\%$ credible intervals decrease together with the noise level $\sigma_\eta = 10^{-1}$ (left) and $\sigma_\eta = 10^{-2}$ (right).}
\label{fig:pendulum_Lasso_data}
\end{figure}
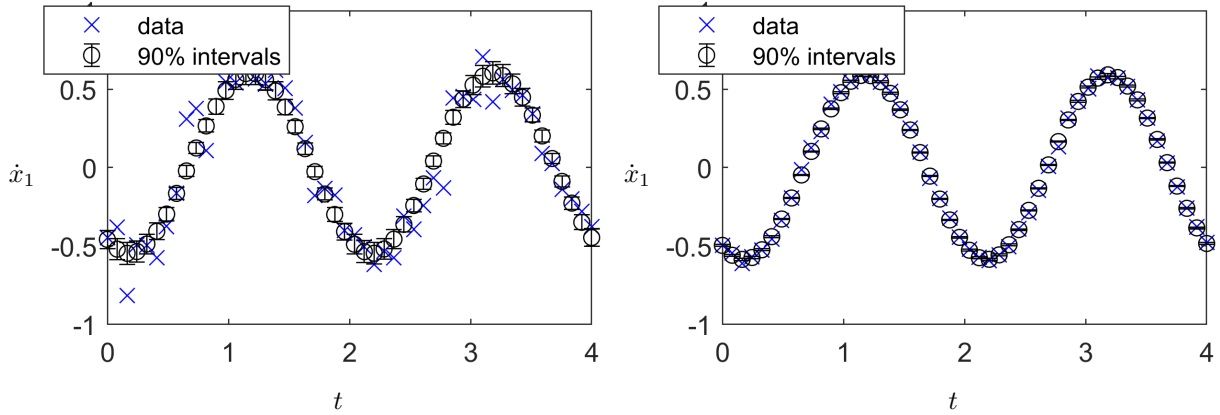
The simple pendulum example, adapted from \cite{galioto2020bayesian}, enables easy testing of the method. 
The dynamical system is given by 
\begin{equation}
    \frac{d}{d t} 
    \left ( 
    \begin{array}{c}
         x_1 \\
         x_2
    \end{array}
    \right )
    = 
    \left(
    \begin{array}{cc}
         0& 1 \\
         -g/L & 0
    \end{array}
    \right )
    \left ( 
    \begin{array}{c}
         x_1 \\
         x_2
    \end{array}
    \right ),
\end{equation}
where $g = 9.81\,$ms$^{-1}$ and $L=1\,$m. 
The system is simulated in time with MATLAB's built-in Runge Kutta ODE45 time integrator over a time interval $[0,4]\,$s. Using this example, we investigate different activation functions (hence, different priors) and the influence of the data noise. 
Data are generated at $m=50$ uniformly spaced time intervals. 
In particular, the data vector is created as 
\[
\vec{z}_j = \dot{\vec{x}}(t_j) + \sigma_\eta \vect{\eta}_j, \quad  j=1,\ldots,m
\]
where $t_j$ denotes the discrete time, $\sigma_\eta$ denotes the data noise amplitude, and $\vect{\eta}_j = (\eta_{j,1},\eta_{j,2})^\top, \eta_{j,1},\eta_{j,2} \sim \mathcal{N}(0,1)$ are independently drawn. The library in this case is chosen as
\[
\Theta(\vec{X}) = \left[\vec{X} \ \vec{X}^2 \right].
\]
The two equations defining the system are identified individually with the neuronized MCMC algorithm. 
The hyperparameters are chosen, as described in Section~\ref{sec:Bsindy}.
For each MCMC run a sample size of $M=10^5$ elements is chosen, which yields a Geweke score
\begin{equation}
    \label{eq:Geweke}
    Z = \frac{\mu_A - \mu_B}{\sqrt{\frac{\sigma_A}{n_A} + \frac{\sigma_B}{n_B}}}
\end{equation}
smaller than $2$.  
In equation \eqref{eq:Geweke}, $\mu_{A/B},\sigma_{A/B}$ refer to the moments of Markov sub-chains of size $n_A = 0.1 M, n_B = 0.5 M$ extracted from the beginning and the end, respectively. 
We always neglect the first 20$\%$ of the Markov Chain to remove the burn-in phase. 

In Figure~\ref{fig:pendulum_Lasso_posterior}, we present the first component of the reference solution along with the identified system with the Lasso prior-based posterior distribution. 
The latter is obtained by constructing a family of dynamical models according to \eqref{eq:SINDy_component}, 
and solving them with the same Runge Kutta method. 
The posterior mean represents the arithmetic mean over all posterior trajectories and the $90 \%$ credible interval bounds cover $90 \%$ of all trajectories. 
We observe the larger uncertainty band, reflected by the posterior credible interval on the left ($\sigma_\eta = 10^{-1}$), compared to the right ($\sigma_\eta = 10^{-2}$). 
Since for the Lasso prior $\alpha_0 = 0$, only a single hyperparameter needs to be selected, which is chosen according to the signal-to-noise ratio as \cite{shin2021neuronized}, 
\begin{equation}
 \label{eq:tau_wExpression}
 \tau_w = \| \vect{\xi}\|_2/(\sqrt{p \mathbb{E}[T^2(\alpha)]}),
\end{equation}

where we draw $\alpha \sim \mathcal{N}(0,1)$ and approximate the expected value with the Monte Carlo method. 
This yields $\tau_w = 0.0039$ and $\tau_w = 0.0148$ for the first and second equation, respectively. The posterior distributions for the Horshoe and discrete spike and slab prior are qualitatively similar.
They are reported in Appendix~\ref{sec:appendix_pendulum}. 
In Figure~\ref{fig:pendulum_Lasso_data}, we report the data $\vec{z}$ and the posterior credible intervals for the linear model $\Theta(\vec{X}) \vect{\xi}$, for each equation separately. 
We again observe a decreasing uncertainty, when the noise level is reduced from $\sigma_\eta=10^{-1}$ on the left, to $\sigma_\eta=10^{-2}$ on the right. 
The results for the other available prior choices are largely similar and are not reported here. 

\begin{figure}[t!]
    \begin{minipage}{0.33\textwidth}
    \centering
    \begin{tikzpicture}
    \node at (0,2) {Lasso};
    \node at (0,0) {\includegraphics{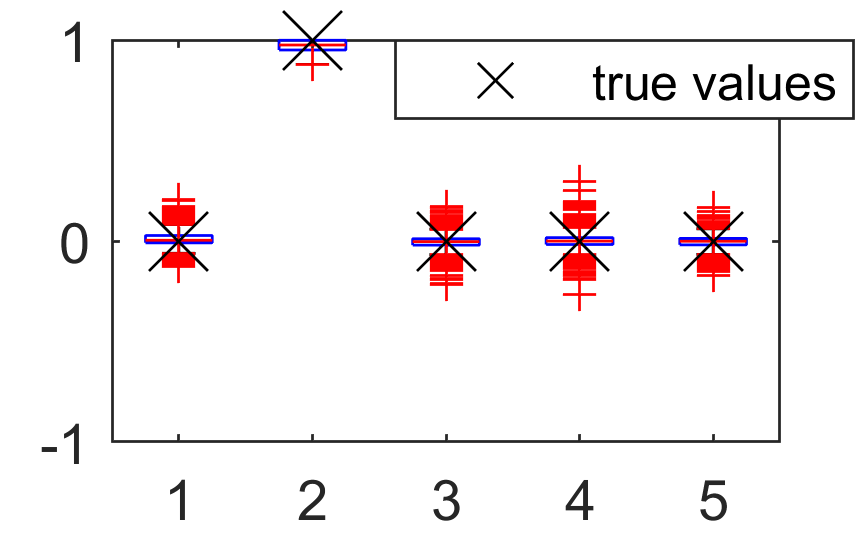}};
    \node at (-3,0) {$\xi_i$};
    \node at (0.1,-2) {$i$};
    \end{tikzpicture}
    \end{minipage}
    \begin{minipage}{0.33\textwidth}
    \centering
    \begin{tikzpicture}
    \node at (0,2) {Horseshoe};
    \node at (0,0) {\includegraphics{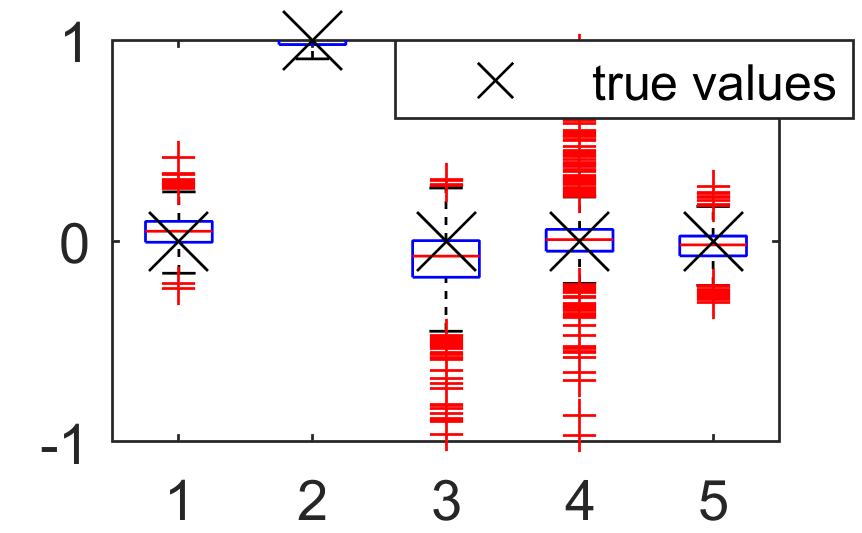}};
    \node at (-3,0) {$\xi_i$};
    \node at (0.1,-2) {$i$};
    \end{tikzpicture}
    \end{minipage}
    \begin{minipage}{0.33\textwidth}
    \centering
    \begin{tikzpicture}
    \node at (0,2) {ReLu};
    \node at (0,0) {\includegraphics{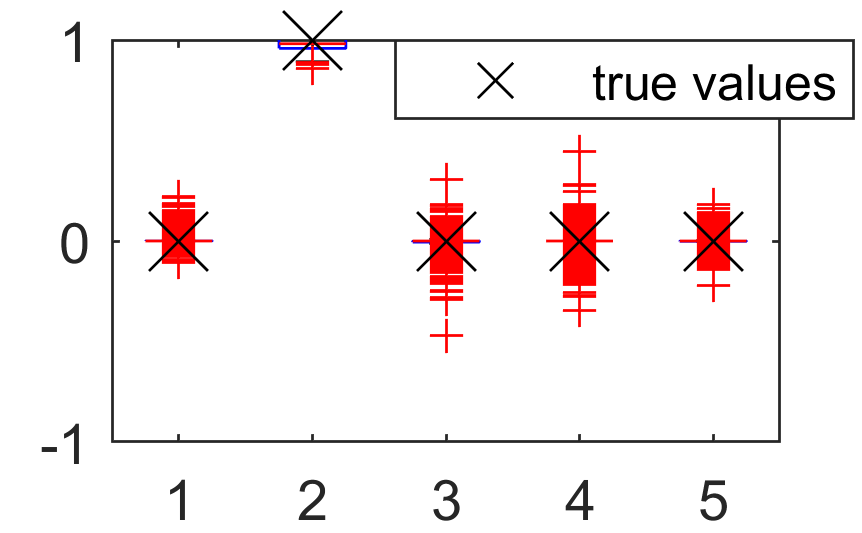}};
    \node at (-3,0) {$\xi_i$};
    \node at (0.1,-2) {$i$};
    \end{tikzpicture}
    \end{minipage}
    \caption{True parameter values for the first equation of the pendulum and box plots for posterior sample of the estimated coefficients. The library basis functions, indexed with $i$, are $x_1$ ($i=1$), $x_2 (i=2)$, $x_1 x_2 (i=3)$, $x_1^2 (i=4)$, $x_2^2 (i=5)$. The results are shown for a noise level $\sigma_\eta=0.1$ and $m=50$ data points are used.}
    \label{fig:Pendulum_Box_Plots}
\end{figure}
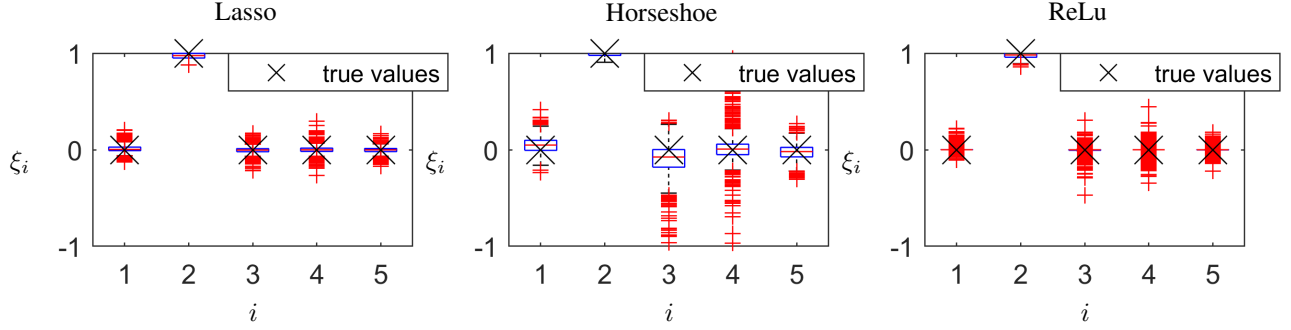

Our investigations are not complete without examining the accuracy of the different prior methods with regard to estimating the regression coefficients. 
In Figure~\ref{fig:Pendulum_Box_Plots}, the true parameter values are shown together with the posterior distributions of the estimated coefficients. 
The posterior distribution is visualized with box plots. 
The Lasso and ReLu prior deliver visually the best estimates, whereas the scatter in the Horseshoe prior is larger. However, comapred to the Lasso, the ReLu prior improves the sparsity in the solution. Whenever there is a real interest in selecting one ``winning'' model, a common choice is selection of the median model, i.e., a basis function is included if $P(\gamma_i = 1 )>0.5$. 
In this regard, both Lasso and Horseshoe priors typically yield the full set of of variables. 
In other words, most coefficients are small but non-zero with probability greater than $0.5$. 
Here, an additional threshold would need to be applied, in order to recover a sparse model. Only the ReLu prior delivers a sparse model directly based on the median model selection criterion, where only the coefficient associated to $x_1 x_2$ is erroneously included if the noise is too large. In view of these findings, we consider ReLu activation functions for the remaining, more challenging, examples to follow. 

~\newpage
\subsection{Lorenz System}
The second application employing the neuronised MCMC algorithm is the Lorenz system. 
Here too, the reference system dynamics are known.
The Lorenz system is particularly chosen because of its chaotic nature.
It is driven by a set of three coupled nonlinear ordinary differential equations (ODEs) given by
\begin{equation}
    \label{eq:Lorenz system}
    \begin{split}
    \dot{x}_1&=c_1(x_2-x_1), \\
    \dot{x}_2&=x_1(c_2 - x_3) -x_2, \\
    \dot{x}_3&=x_1x_2-c_3 x_3.
    \end{split}
\end{equation}
The selected parameters are $c_1=10$, $c_2=28$ and $c_3=2.667$ with initial conditions $(x_1^0, x_2^0, x_3^0) =(-8,8,27)$.
The simulation is performed with a time step $\Delta t = 0.1$ for a total time of $T=100$. The initial state $(x_1^0, x_2^0, x_3^0)^\top$ is propagated in time using MATLAB's ODE45 function, and the resulting time series $[\vec{x_1} \vec{x_2} \vec{x_3}]$ is used to compute the derivatives $\vec{\dot{x}_1}, \vec{\dot{x}_2}$ and $\vec{\dot{x}_3}$. 
To simulate numerical differentiation noise, normally distributed random noise $\boldsymbol{\eta}\sim  \mathcal{N}(\boldsymbol{0}_{m\times1},\vec{I}_{m\times m})$ is separately added to $\vec{\dot{x}_1}, \vec{\dot{x}_2}$ and $\vec{\dot{x}_3}$. 

\begin{figure}[ht]
\begin{minipage}[b]{0.33\linewidth}
\centering
\includegraphics[width=2.25in, height=1.7in]{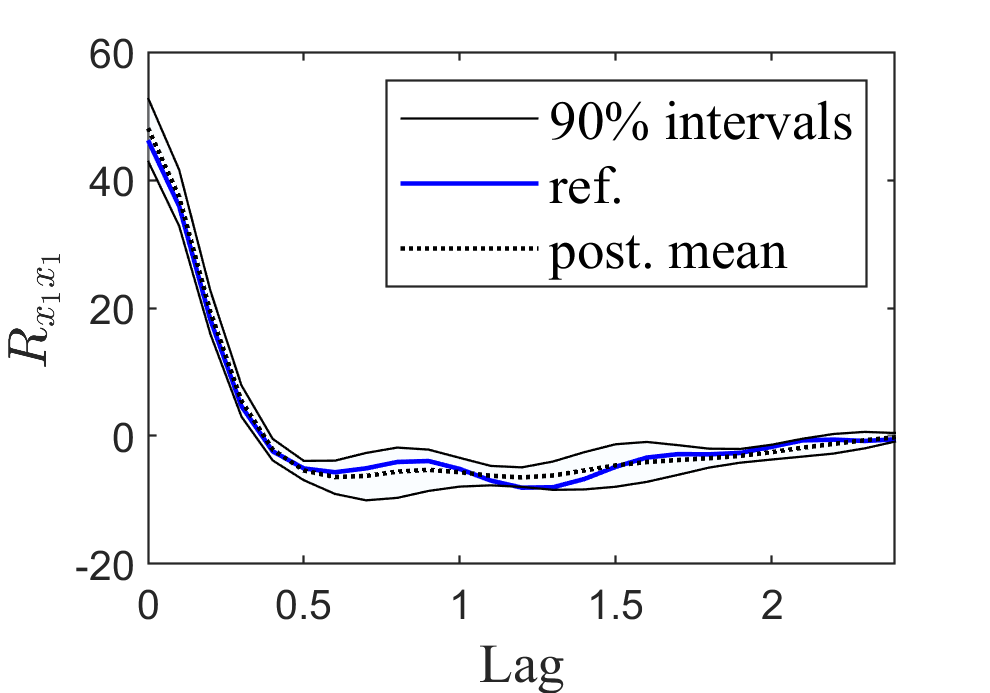}
\end{minipage}
\hspace{0.2cm}
\begin{minipage}[b]{0.33\linewidth}
\centering
\includegraphics[width=2.25in, height=1.7in]{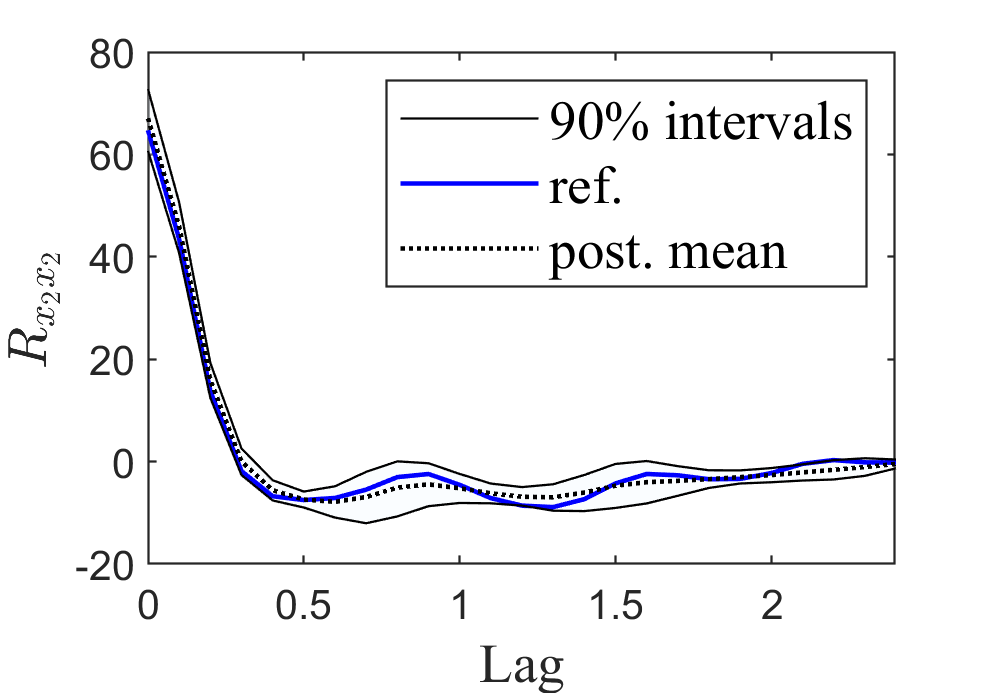}
\end{minipage}
\hspace{0.2cm}
\begin{minipage}[b]{0.33\linewidth}
\centering
\includegraphics[width=2.25in, height=1.7in]{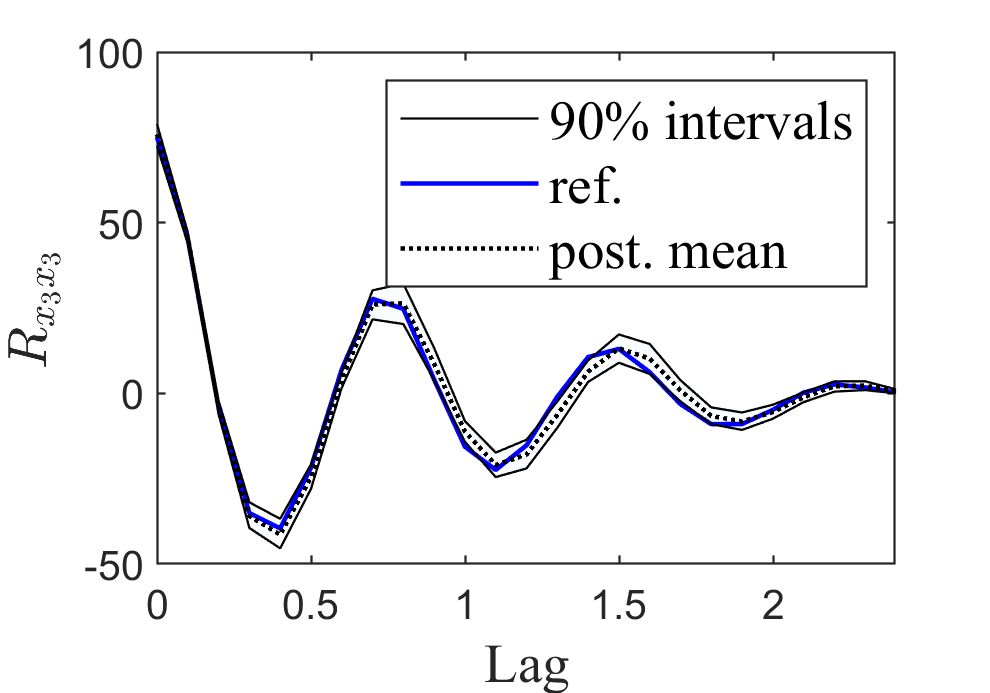}
\end{minipage}
\caption{The reference trajectory, posterior mean and credible intervals covering 90\% of the posterior samples of the autocorrelation distributions $\vec{x}_1,\vec{x}_2$ and $\vec{x}_3$. }
\label{fig:AutoCorrLorrenz}
\end{figure}

\begin{figure}[ht]
\centerline{
\includegraphics[width=1.1\linewidth]{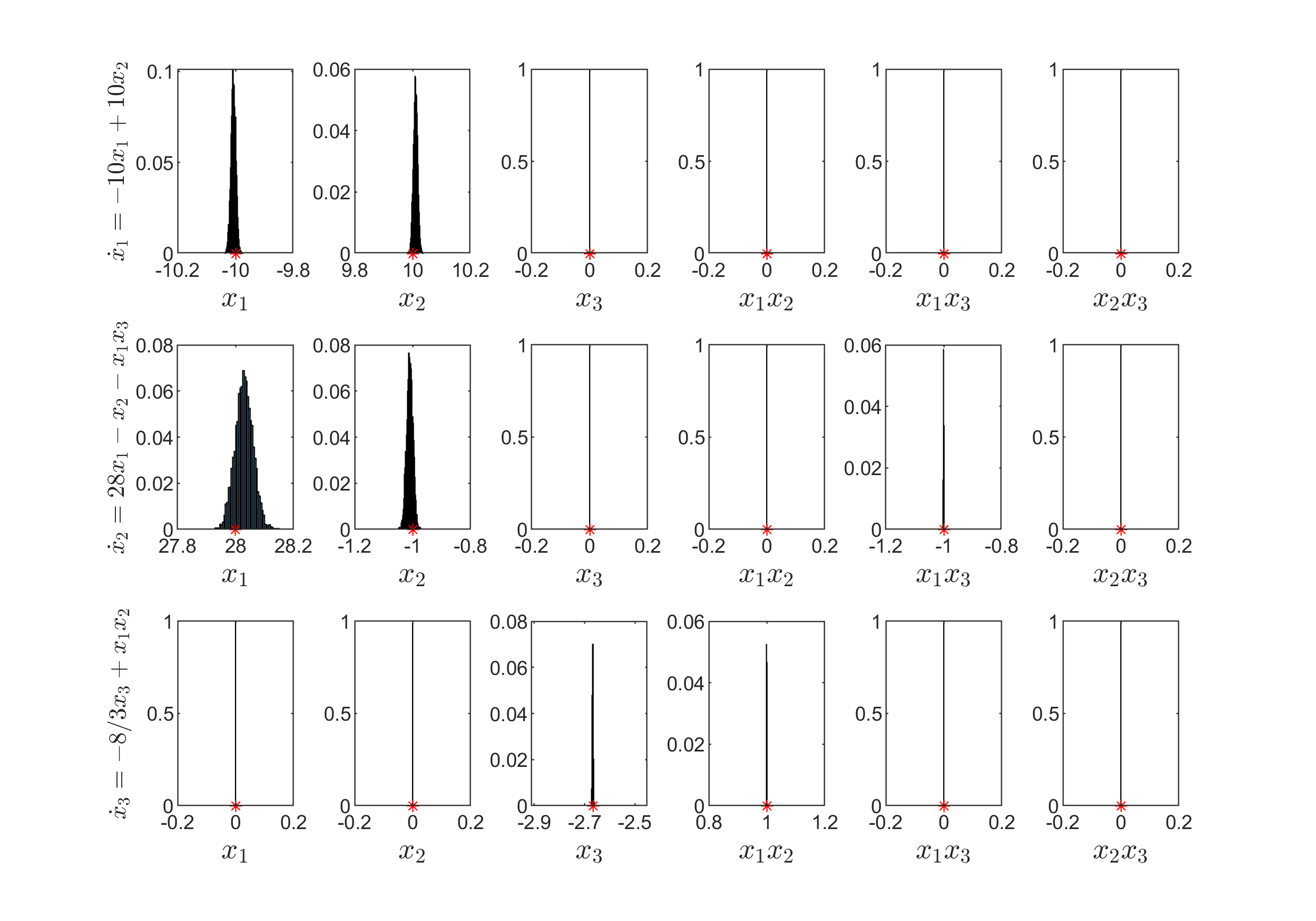}}
\caption{Each row in the above figure corresponds to one particular equation of the Lorenz system. Each row contains the histograms of the coefficients of each of the basis functions comprising $\vec{D}$. The values of the coefficients used to generate the data vectors are indicated with a red star.}
\label{fig:LorenzHisto}
\end{figure}

With the data vectors computed, the neuronised MCMC algorithm is applied once on each of the data vectors. 
For all the three vectors, a discrete spike and slab/ReLu prior with $\alpha_0=0.5$ is used. 
As per equation \eqref{eq:tau_wExpression}, $\tau_w$ the parameter that determines the variance of the prior's slab part, is set to 12.462, 24.791 and 2.518, for the three components $\vec{\dot{x}}, \vec{\dot{y}}$, and $\vec{\dot{z}}$, respectively,
whereas $\alpha_0$ is set to 0.5 to obtain a high degree of sparsity. 
While computing $\tau_w$'s, the least squares estimate $\boldsymbol{\xi}^i=(\vec{D}^{\top}\vec{D})^{-1}\vec{D}^{\top}\vec{z}^i$ is used, 
with $i=1\dots 3$ for the 3 Lorenz equations. Note that the $\tau_w$ values are set before starting the MCMC for each of the equations.
In this example and the following one, the error variance $\sigma_{\eta}^2$ was also updated within the MCMC, by imposing an inverse gamma prior Inv-Gam($a_0, b_0$) on $\sigma_{\eta}^2$, with both $a_0$ and $b_0$ assigned a value of 1. On the imposition of such a prior, the  posterior of $\sigma_{\eta}^2$ too has an inverse gamma form (see \cite{shin2021neuronized}).  
In order to facilitate efficient sampling of $\boldsymbol{\alpha}$ for the case of the ReLu prior, $\boldsymbol{\alpha}$ is directly sampled from a mixture of two truncated Gaussians \cite{shin2021neuronized}, instead of the Random Walk Metropolis Hastings step in Algorithm \ref{alg:BSindy}.
This sampling strategy is performed for both the current and the next application.
The following basis library: $\vec{D}(\vec{x_1},\vec{x_2},\vec{x_3}) = \left[\vec{x_1}, \vec{x_2} , \vec{x_3} , \vec{x_1x_2} , \vec{x_1x_3} , \vec{x_2x_3} \right]$ is employed.
The MCMC is run for a total of $M=10000$ iterations and the first 3000 sample points are discarded as burn-in sample points. 
This yields a Geweke score smaller than 2.

The posterior distributions of the coefficients $\boldsymbol{\xi}$ are presented in Figure \ref{fig:LorenzHisto}. 
Each row corresponds to one equation of the Lorenz system. 
The reference coefficients are marked with red dots.
As the figure shows, the histograms of coefficients corresponding to insignificant bases are associated  with spikes at zero,
whereas those of the non-zero coefficients are correctly centered around their corresponding reference values. With the obtained posteriors, using a median model selection criterion, the correct basis functions for each of the equations in equation \eqref{eq:Lorenz system} are recovered. In Table~\ref{tab:Single point estimates}, we compare the single point estimates obtained using the neuronised prior algorithm and the SINDy algorithm against the reference values. For the neuronised prior estimates, the posterior mean of the coefficients included in the median model is used. For SINDy, the estimates obtained for $\lambda=0.1$ are shown. It is to be noted that the SINDy estimates for other values of $\lambda$, that is $\lambda=0.3, 0.5, 0.7$ and 0.9, were also computed and they were found to be identical to the estimate for $\lambda=0.1$, and hence they are not reported here. It can be seen that both the SINDy and the neuronised prior estimates match very closely with the reference values. 

\begin{table}[]
\begin{center}
\begin{tabular}{|c|c|c|c||c|c|c||c|c|c|}
\hline
\multicolumn{4}{ |c|| }{$\dot{x}_1=10(x_2-x_1)$} &
\multicolumn{3}{ |c|| }{$\dot{x}_2=x_1(28-x_3)-x_2$} & \multicolumn{3}{ |c| }{$\dot{x}_3=x_1x_2 -8/3x_3$} \\ \hline 
\multirow{2}{*}{Bases} &\multirow{2}{*}{Ref.} & \multirow{2}{*}{Neu. Pr.} & {SINDy}  &\multirow{2}{*}{Ref.} & \multirow{2}{*}{Neu. Pr.} & {SINDy} &\multirow{2}{*}{Ref.} & \multirow{2}{*}{Neu. Pr.} & {SINDy}\\  & & & $\lambda=0.1$ & & & $\lambda=0.1$ & & & $\lambda=0.1$ \\ \hline
$x_1$ & -10 &  -10.0076 & -10.0076 & 28 &  28.0295 & 28.0296 & 0 &  0 & 0  \\ \hline
$x_2$ & 10 &  10.0101 & 10.0102 & -1 & -1.0097   & -1.0097 & 0 &  0 & 0  \\ \hline
$x_3$ & 0 &  0 & 0 & 0 & 0 & 0 & -2.6667 &  -2.6679 & -2.6679  \\ \hline
$x_1x_2$ & 0 &  0 & 0 & 0 & 0 & 0 &1 &  0.9999 & 0.9999  \\ \hline
$x_1x_3$ & 0 &  0 & 0 & -1 &  -1.0007 & -1.0007 & 0 & 0 & 0 \\ \hline
$x_2x_3$ & 0 &  0 & 0 & 0 & 0 & 0 & 0 & 0 & 0  \\ \hline
\end{tabular}
\end{center}
\bigskip
\caption{Single point estimates comparison between SINDy and neuronized prior system dynamics identification. For the latter the posterior means of the median model are shown.}
\label{tab:Single point estimates}
\end{table}

With the posteriors of $\boldsymbol{\vec{\xi}}^i$ determined, the states $\vec{x}_1, \vec{x}_2$ and $\vec{x}_3$ are propagated through equation \eqref{eq:Lorenz system}. 
Since a direct comparison of time series is meaningless for chaotic systems, a quantitative assessment of the model requires an alternative metric.
In this study, we compare the autocorrelation function, computed as $R_{\vec{x}\vec{x}}(t_1,t_2)=\mathbb{E}[\vec{x}_{t_1}\vec{x}_{t_2}]$, of the reference data and that of the model.
Figure \ref{fig:AutoCorrLorrenz} presents window-averaged auto-correlation distributions for the states $\vec{x}_1$, $\vec{x}_2$ and $\vec{x}_3$. Forty time-windows of length 2.5 units are used for the averaging. It can be observed that the reference curves match the posterior mean distributions very well. 

\subsection{Aerodynamic Application}
The final application employs measured data from an aerodynamic experiment.
The experiment consists of an airfoil equipped with a pitching flap.
Here, we are interested in identifying a model for the unsteady lift coefficient $C_l$ as a function of the pitch angle $\delta$ and enriched features thereof. 
Both $C_l$ and $\delta$ are experimentally acquired in real-time with surface pressure sensors and position encoders, respectively.
The particular test case is purposely selected to exhibit challenging dynamics through dynamic stall.
Details on the experimental setup as well as the aerodynamic problem are found in \cite{pohl_quantification_2020} and \cite{pohl_dynamic_2019}.

The dynamics is learned from the following basis library $$\vec{D}(\vec{C}_l,\boldsymbol{\delta})=[\vec{C}_l,\dot{\boldsymbol{\delta}},\boldsymbol{\delta},\boldsymbol{\delta}\dot{\boldsymbol{\delta}},\dot{\boldsymbol{\delta}}^2,\boldsymbol{\delta}^2,\dot{\boldsymbol{\delta}}^3,\boldsymbol{\delta}^3,\dot{\boldsymbol{\delta}}_{\frac{1}{4}},\boldsymbol{\delta}_{\frac{1}{4}},\dot{\boldsymbol{\delta}}_{\frac{1}{2}},\boldsymbol{\delta}_{\frac{1}{2}},\dot{\boldsymbol{\delta}}_{\frac{1}{8}},\boldsymbol{\delta}_{\frac{1}{8}}],$$ consisting of 14 basis functions, where $\boldsymbol{\delta}\dot{\boldsymbol{\delta}}$ refers to an element-wise multiplication of both vectors. 
Here, the subscripts $\frac{1}{8}$, $\frac{1}{4}$ and $\frac{1}{2}$ refer to corresponding time-delayed basis functions by fractions of a time period. The derivatives $\dot{\vec{C}}_l$ and $\dot{\boldsymbol{\delta}}$ are obtained by first smoothing $\vec{C}_l$ and $\boldsymbol{\delta}$ and then using MATLAB's gradient function. A noise
vector $\boldsymbol{\eta}\sim \mathcal{N}(\boldsymbol{0}_{m\times1},0.01^2 \ \vec{I}_{m\times m})$ is added to $\dot{\vec{C}}_l$ to obtain $\vec{z}$ for the subsequent steps.  
The model coefficients $\boldsymbol{\xi}$ are estimated with the neuronised MCMC algorithm employing a discrete spike and slab/ReLu prior with $\alpha_0=1.5$ and $\tau_w=227.468$.
The hyper-parameter $\alpha_0$ is assigned this value to obtain a high degree of sparsity, and the parameter $\tau_w$ is set as per equation \eqref{eq:tau_wExpression}. 
As described in the Lorenz example, $\tau_w$ is set before starting the MCMC and the least squares estimate of $\boldsymbol{\xi}$ is used in its computation.
Since we are mainly interested in the unsteady lift dynamics, both $\vec{D}$ and the data vector are first de-meaned.
To simplify the sparse optimisation procedure, we additionally normalize $\vec{D}$ columnwise with the respective standard deviations of each basis vector. 
Similar to the Lorenz example, the MCMC is run for $M=10000$ iterations, which results in a Geweke score of less than 2. The first 3000 iterations are discarded as burn-in sample points. The magnitudes of $a_0$ and $b_0$ pertaining to the inverse gamma prior for $\sigma^2$ are set to 1, whereas the initial value of $\sigma$ is set to 0.1. The value of $\sigma$ gets updated as part of the MCMC procedure.  

\begin{figure}[ht]
 \centerline
 {\includegraphics[width=1\linewidth]{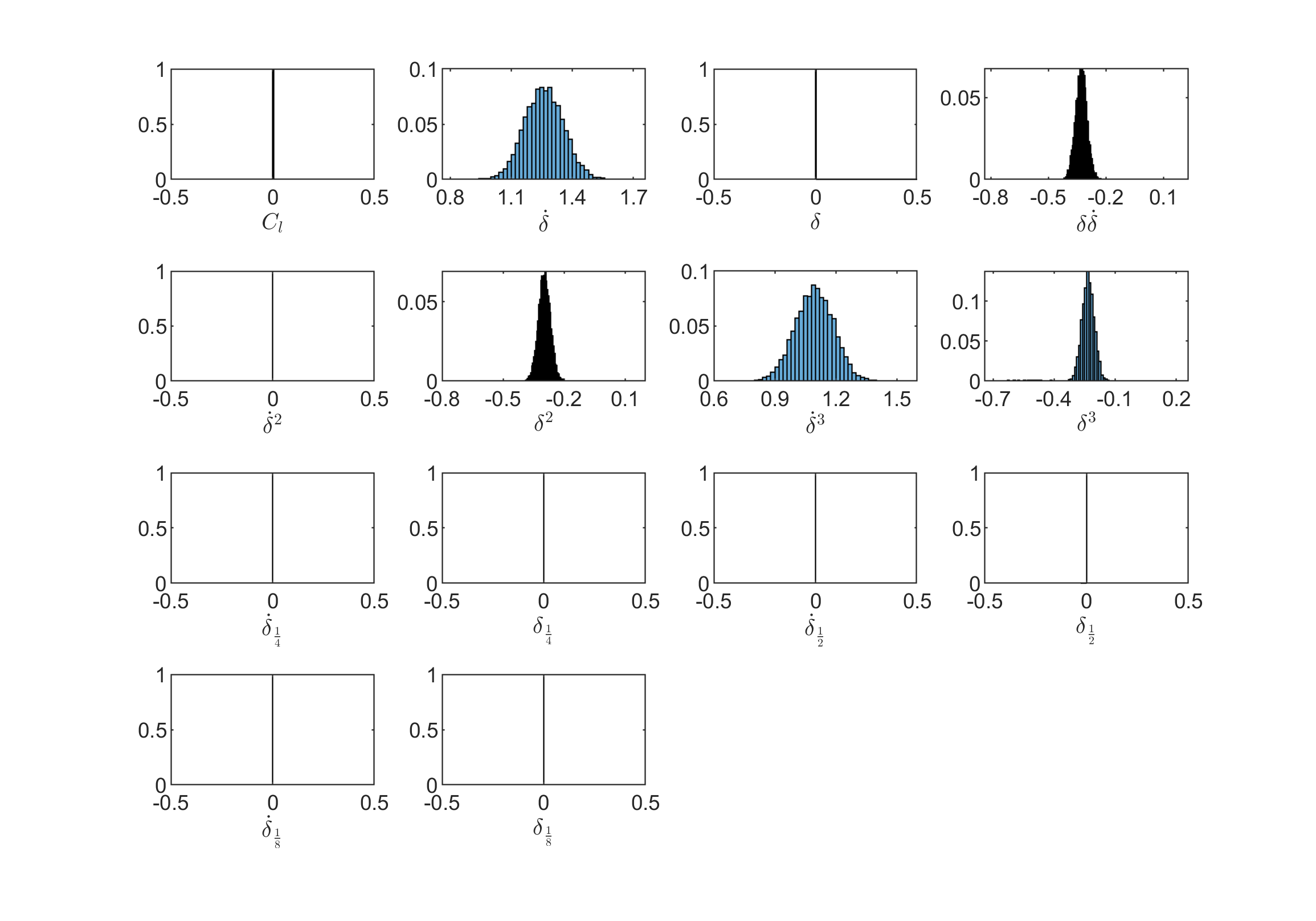}}
 \caption{Posterior distributions of the model coefficients $\boldsymbol{\xi}$ for the aerodynamic example. 
 Starting with a library of 14 basis functions, the algorithm identifies five bases $\dot{\boldsymbol{\delta}},\boldsymbol{\delta}\dot{\boldsymbol{\delta}},\boldsymbol{\delta}^2,\dot{\boldsymbol{\delta}}^3$ and $\boldsymbol{\delta}^3$ that contribute to the lift dynamics.}
 \label{fig:AeroHistograms} 
 \end{figure}

The resulting posterior distributions of the coefficients $\boldsymbol{\xi}$ are shown in Figure \ref{fig:AeroHistograms}. 
As the figure shows, if we consider the median model, only five basis functions contribute to the dynamics, $\dot{\boldsymbol{\delta}},\boldsymbol{\delta}\dot{\boldsymbol{\delta}},\boldsymbol{\delta}^2,\dot{\boldsymbol{\delta}}^3$, and $\boldsymbol{\delta}^3$.
This yields a model with the following structure,
\begin{equation}
\label{eq:AeroEqn}
\dot{\vec{C}_l} \approx \xi_1 \dot{\boldsymbol{\delta}} + \xi_2 \boldsymbol{\delta}\dot{\boldsymbol{\delta}}  + \xi_3 \boldsymbol{\delta}^2 + \xi_4 \dot{\boldsymbol{\delta}}^3 + \xi_5 \boldsymbol{\delta}^3
\end{equation}
where the coefficients can assume values according to their respective posteriors. The identified model offers an intriguing structure. While the pitch angle rate $\dot{\boldsymbol{\delta}}$ is an expected basis in unsteady aerodynamics modeling, the remaining ones are not.
Their role is to model nonlinear interactions deviating from harmonic input-output response. 
This nonlinearity is caused by the so-called dynamic stall, where the rapid pitching causes a strong vortex to be shed from the leading edge that travels downstream over the wing.
Hence, these basis functions are compensating for the assumed restrictive model structure.
Airfoils with dynamic stall require more complex dynamical models, such as a Duhamel integral for the indical response approach \cite{herbert_wagner_1925}, or an additional state variable for the ONERA modeling approach \cite{mcalister_application_1984}. 

\begin{figure}[h!]
  \centerline
 {\includegraphics[width=1\linewidth]{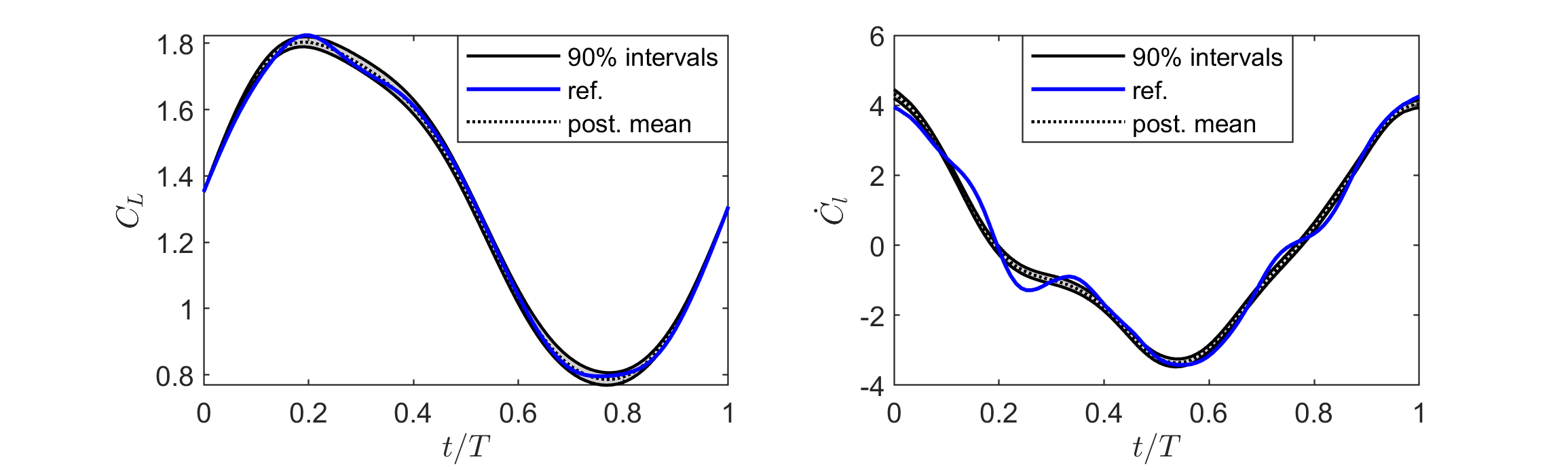}}
 \caption{(Left) the lift coefficient $C_l$ and (right) its time derivative $\dot{C_l}$ over one pitching period $T$. 
 The three distributions are the reference trajectory, posterior mean, and credible intervals covering 90\% of the posterior samples.
} 
 \label{fig:Derivative&State}
 \end{figure}

As done in the previous examples, the entire posterior of $\boldsymbol{\xi}$ is used to reconstruct the dynamics $\dot{\vec{C}_l} \approx \vec{D}\boldsymbol{\xi}$, which is subsequently used to reconstruct $\vec{C}_l$. 
The reference lift coefficient distribution along with the distribution of its time derivative over a time period of one pitching period $T$ are shown in Figure \ref{fig:Derivative&State}. 
Despite the restrictive model structure, both distributions compare favorably to the reference data.

\section{Summary and conclusions}
\label{sec:Summary}
In this paper, we have presented an uncertainty-aware method for sparse dynamical system identification and applied it to both generic model examples and an aerodynamic application with real data. Here, the main contribution to Bayesian dynamical system identification was the adaption of a unified prior family, the neuronized prior. In particular, by choosing different activation functions we were able to recover and implement well-known sparsity priors, such as spike and slab, Horseshoe and Lasso priors in particular, in a common framework. We then employed dedicated MCMC methods to infer dynamical reduced order models together with parameter and structural model uncertainties. The framework also allows to obtain a single model with a model selection strategy, which is desired from an application point of view to gain physical insights from the data. Several directions are to be further explored. We have limited the considerations to moderate basis dimensions and additional efforts are needed to address large-scale problems. Moreover, including model parameters into the framework would allow to answer relevant tasks in control applications. Finally, more general noise structure should be considered to reflect the structure of complex data.

\bibliographystyle{unsrt}
\bibliography{references.bib}

\begin{appendices}

 \section{Additional Results for the Pendulum}
 \label{sec:appendix_pendulum}
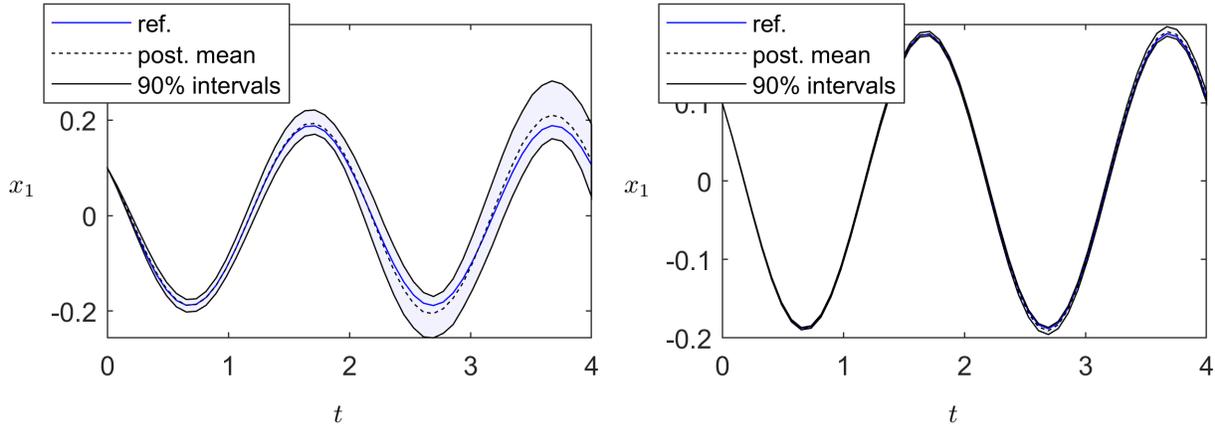
\begin{figure}[h!]
\begin{minipage}{0.49\textwidth}
\centering
\begin{tikzpicture}
\node at (0,0) {\includegraphics{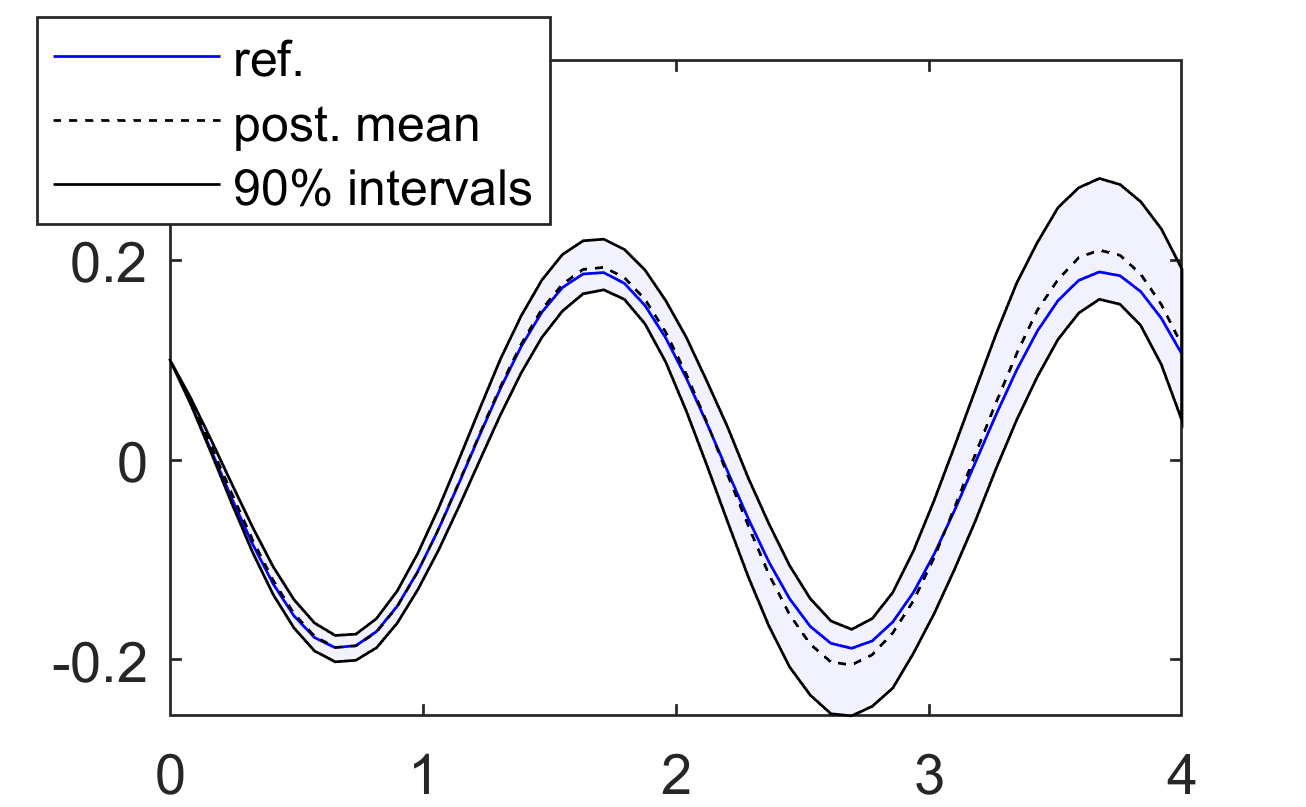}};
\node at (-4.2,0) {$x_1$};
\node at (0,-3) {$t$};
\end{tikzpicture}
\end{minipage}
\begin{minipage}{0.49\textwidth}
\centering
\begin{tikzpicture}
\node at (0,0) {\includegraphics{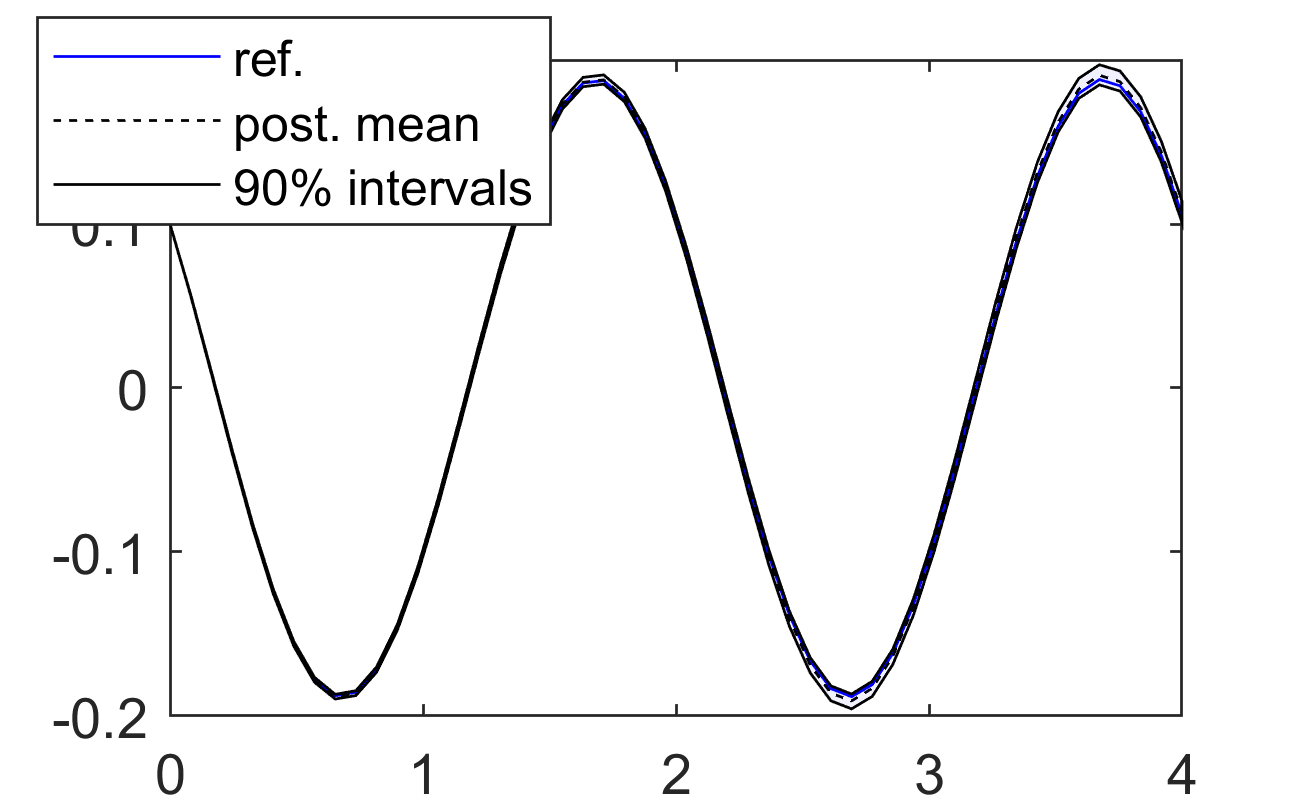}};
\node at (-4.2,0) {$x_1$};
\node at (0,-3) {$t$};
\end{tikzpicture}
\end{minipage}
\caption{Results for the identification of the linear pendulum with a Horseshoe prior. Reference trajectory, posterior mean and credible intervals covering $90 \%$ of the posterior samples. These samples are obtained by computing the time trajectory for posterior samples of $\vect{\xi}$. Left: data size $n=10$ and noise level $\sigma_\eta = 10^{-1}$. Right: data size $n=50$ and noise level $\sigma_\eta = 10^{-2}$.}
\label{fig:pendulum_Horseshoe_posterior}
\end{figure}

\begin{figure}[h!]
\begin{minipage}{0.49\textwidth}
\centering
\begin{tikzpicture}
\node at (0,0) {\includegraphics{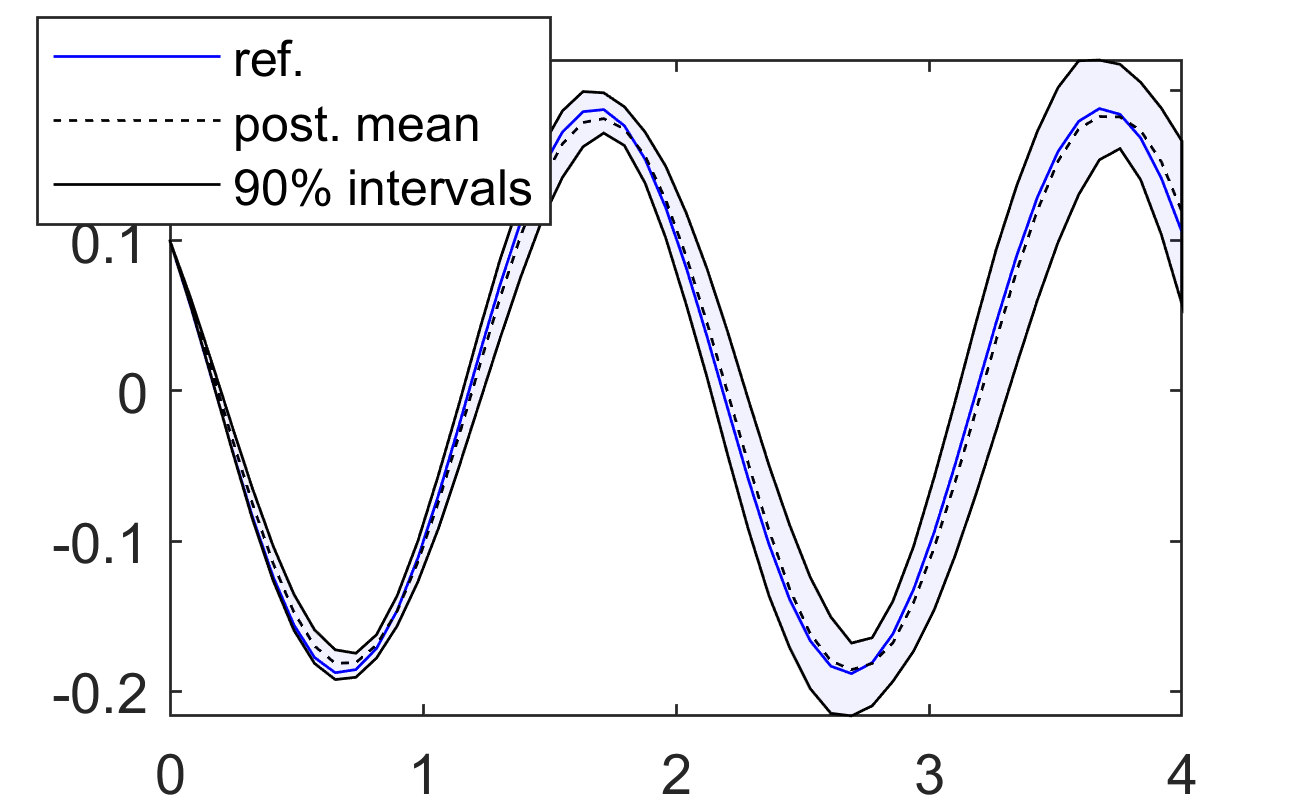}};
\node at (-4.2,0) {$x_1$};
\node at (0,-3) {$t$};
\end{tikzpicture}
\end{minipage}
\begin{minipage}{0.49\textwidth}
\centering
\begin{tikzpicture}
\node at (0,0) {\includegraphics{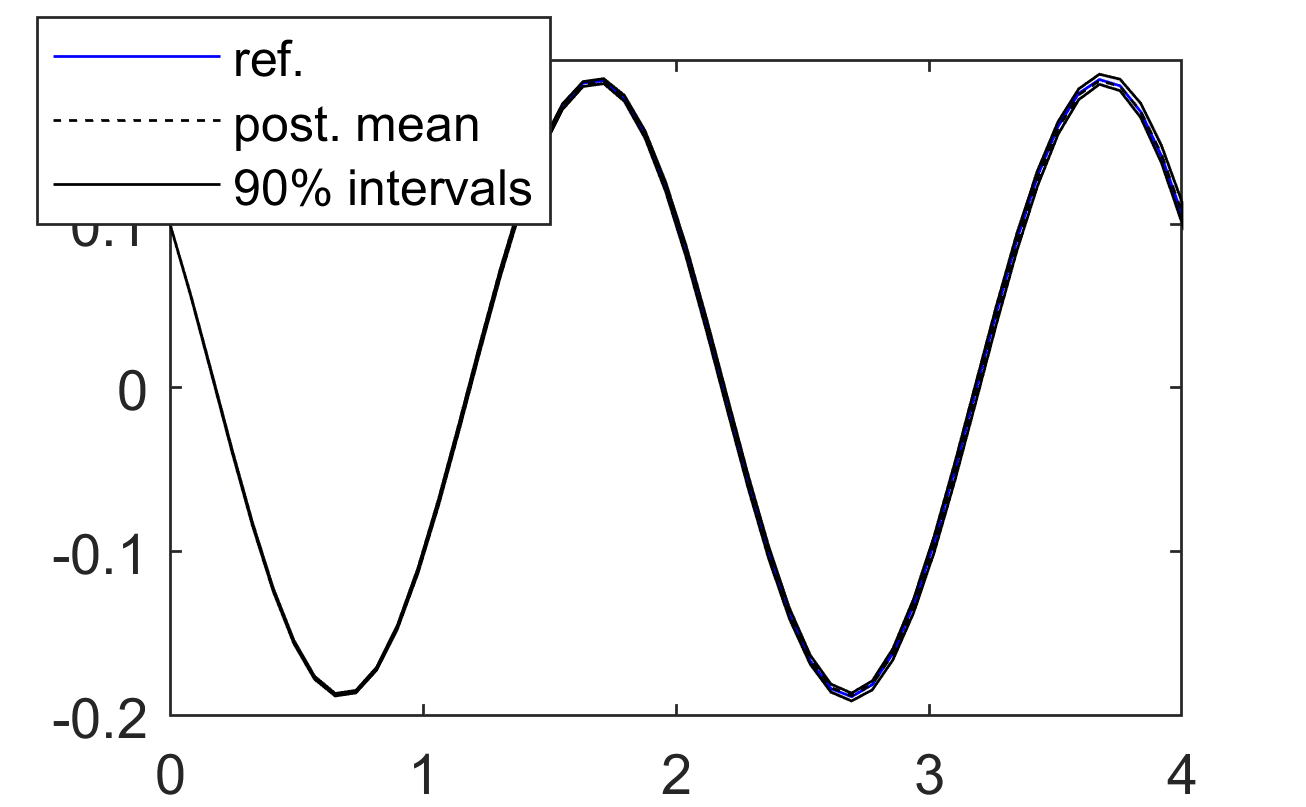}};
\node at (-4.2,0) {$x_1$};
\node at (0,-3) {$t$};
\end{tikzpicture}
\end{minipage}
\caption{Results for the identification of the linear pendulum with a discrete spike and slab prior. Reference trajectory, posterior mean and credible intervals covering $90 \%$ of the posterior samples. These samples are obtained by computing the time trajectory for posterior samples of $\vect{\xi}$. Left: data size $n=10$ and noise level $\sigma_\eta = 10^{-1}$. Right: data size $n=50$ and noise level $\sigma_\eta = 10^{-2}$.}
\label{fig:pendulum_ReLu_posterior}
\end{figure}
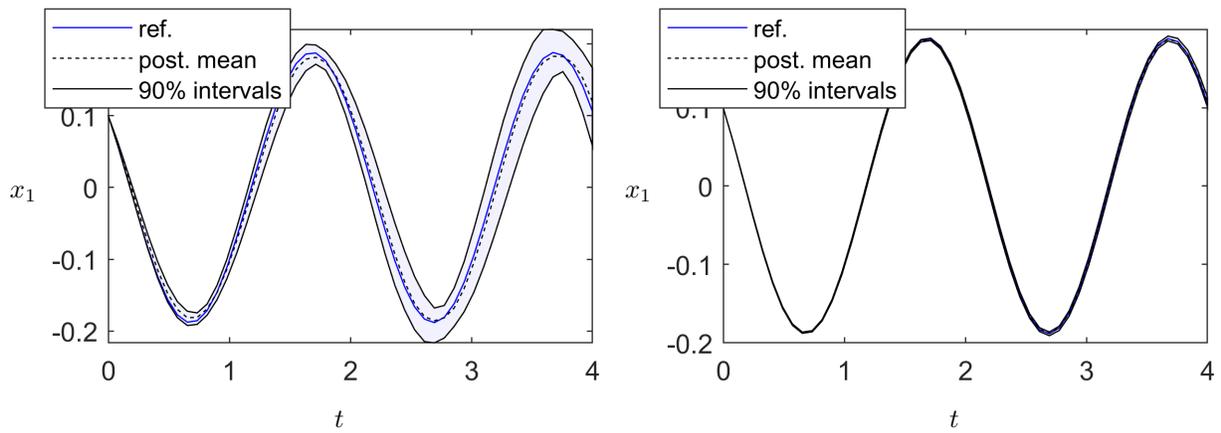

 In Figure~\ref{fig:pendulum_Horseshoe_posterior} we report the posterior results for the Horseshoe prior with activation function $T(x) = \exp(0.5 \mathrm{sign}(x) x^2 + 0.733 x)$. Again $\alpha_0 = 0$, however, a different value for $\tau_w$ needs to be chosen to yield good results. In \cite{shin2021neuronized}, it is recommended to choose $\tau_w$ such that $1-\mathbb{E}[1/(1+T(\alpha)^2 \tau_w^2)]$, $\alpha \sim \mathcal{N}(0,1)$ equals the expected portion of zero values, which is $\approx 0.2$ in our case. In this regard, a value $\tau_w$ seems appropriate and indeed yields satisfactory results for the case $\sigma_{\eta}=0.1$. However, for $\sigma_\eta=0.2$ we observed that larger values are needed. The results reported in Figure~\ref{fig:pendulum_Horseshoe_posterior} (right) have been obtained with $\tau_w =10,20$ for the first and second equation respectively. 
 Finally, in Figure~\ref{fig:pendulum_ReLu_posterior} the results for the ReLu activation function are reported, which corresponds to a discrete spike and slab prior. Here, the hyperparameter $\tau_w$ is again chosen according to the signal-to-noise ratio, as described in Section~\ref{sec:pendulum}, where also the concrete numerical values are given. In case of the ReLu activation function, a second hyperparameter $\alpha_0$ has to be chosen. In \cite{shin2021neuronized} this parameter was related to the expected fraction of non-zero coefficients $\eta$ as $\alpha_0= \Phi^{-1}(\eta)$, where $\Phi$ refers to the CDF of the standard normal distribution. Note, that the parameter $\eta$ is commonly employed as the Bernoulli-hyperparameter in the spike and slab formulation with a latent, model-inclusion, variable. 

\end{appendices}

\end{document}